\documentclass[letterpaper]{article} 
\usepackage{aaai2027} 
\usepackage[hyphens]{url} 
\usepackage{graphicx} 
\usepackage{natbib} 
\usepackage{caption} 
\usepackage{amsmath}
\usepackage{amssymb}
\usepackage{algorithm}
\usepackage{algorithmic}
\usepackage{booktabs}
\usepackage{xcolor}
\usepackage[table]{xcolor}
\usepackage{enumitem}
\usepackage{subcaption}
\usepackage{siunitx}
\usepackage{tabularx}
\usepackage[most]{tcolorbox}
\usepackage{etoolbox}
\usepackage{multicol}
\usepackage{caption}

\usepackage{newfloat}
\usepackage{listings}
\DeclareCaptionStyle{ruled}{labelfont=normalfont,labelsep=colon,strut=off} 
\floatstyle{ruled}
\newfloat{listing}{tb}{lst}{}
\floatname{listing}{Listing}

\title{PACE: Primitive-Aware Code Evolution for Automated Algorithm Design}
\nocopyright
\author{
    Zhuoliang Xie\textsuperscript{\rm 1},
    Ruihao Zheng\textsuperscript{\rm 1},
    Xiang Xu\textsuperscript{\rm 1},
    Genghui Li\textsuperscript{\rm 2},
    Zhengkun Wang\textsuperscript{\rm 1,}\corresponding
}
\affiliations{
    \textsuperscript{\rm 1}Southern University of Science and Technology\\
    \textsuperscript{\rm 2}Shenzhen University\\
    \{xiezl2025, zhengrh2024\}@sustech.edu.cn, xiangxu5-c@my.cityu.edu.hk, \\ligh@szu.edu.cn, wangzhenkun90@gmail.com
}

\begin{document}

\maketitle

\begin{abstract}
Large Language Model (LLM)-based automated algorithm design typically evolves algorithms as complete, indivisible programs. While this whole-program perspective simplifies the search space, it fundamentally couples the useful local logic to its host program. Consequently, valuable code snippets vanish when the overall program is discarded, making it highly difficult to assess the  contribution of individual algorithmic components.
To address this, we propose Primitive-Aware Code Evolution (PACE), which decouples local logic from complete programs by representing it as persistent units called Executable Algorithmic Primitives (EAPs). To enable code-level transfer, PACE maintains a dynamic set of EAPs. Algorithm evolution is driven by primitive-aware operators that structurally guarantee the retention and cross-program transfer of these components. To evaluate them effectively, PACE leverages Thompson sampling based on parent-relative performance improvements, guiding primitive selection from the set without requiring extra evaluation datasets.
Experiments on four tasks demonstrate that PACE effectively discovers competitive algorithms while structurally preserving valuable algorithmic components.
\end{abstract}

\section{Introduction}


Large Language Models (LLMs) have made executable programs a practical search representation for automated algorithm design (AAD) \cite{liu2024survey}. FunSearch and EoH construct an automated closed loop that integrates LLM-based code generation with program evaluators, thereby effectively improving algorithms without reliance on domain expertise \cite{romeraparedes2024funsearch,liu2024eoh}. Their promising results have driven the development of sophisticated mechanisms, such as algorithmic reflection, tree-based algorithmic exploration, and population management \cite{ye2024reevo,zheng2025mctsahd,dat2025hsevo}.

\begin{figure}[t]
    \centering
    \begin{subfigure}{\linewidth}
        \includegraphics[width=\linewidth]{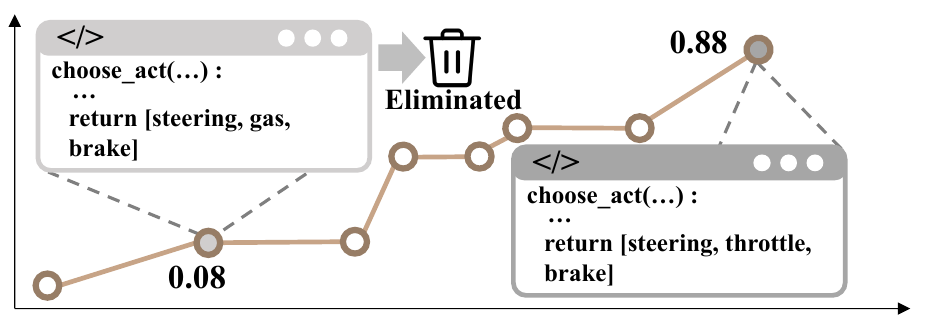}
        \caption{Existing LLM-based AAD methods.} 
        \label{fig:motivation_a}
    \end{subfigure}
    
    \vspace{-2pt} 
    
    \begin{subfigure}{\linewidth}
        \includegraphics[width=\linewidth]{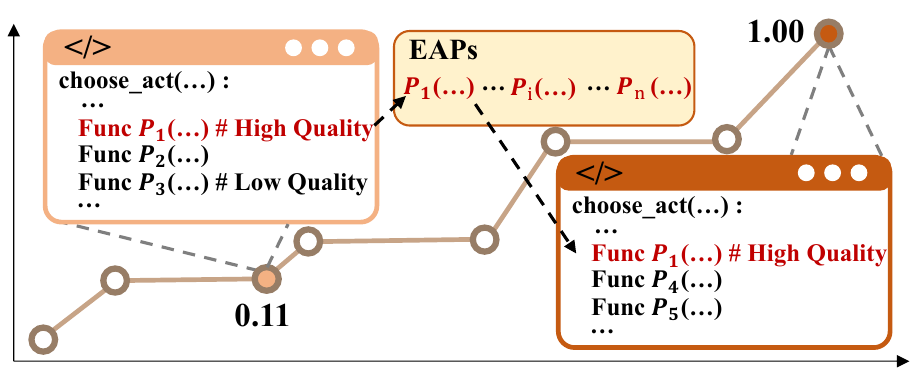}
        \caption{Primitive-aware AAD method (Ours).} 
        \label{fig:motivation_b}
    \end{subfigure}
    
    \caption{(a) Existing AAD methods evaluate whole algorithms. A low-scoring algorithm and its local logic are removed together. (b) PACE retains a useful local logic as a function, a later algorithm can call the same function, thus abstains a better score, even when the original algorithm was removed.}
    \label{fig:motivation}
\end{figure}

Existing AAD methods typically treat the designed algorithm as the minimal unit. Specifically, variation operators such as crossover and mutation \cite{liu2024eoh} place entire parent programs within the LLM context, allowing arbitrary edits to the algorithm without explicit boundary constraints. 
Such a coarse-grained perspective limits the LLM's ability to accurately distinguish between useful and harmful local logics within the algorithm. As shown in \figurename\ \ref{fig:motivation_a}, eliminating a low-performing algorithm entails discarding all of its local logic, including components that could be valuable in a different algorithm. As a result, useful logic may be repeatedly discarded and rediscovered, wasting the search budget and limiting the performance of AAD methods.
Recent progress has also recognized the limitations of directly editing a complete algorithm \cite{yuksel2025evolattice}.


We argue that the AAD process should explicitly preserve useful logic or components for subsequently generated algorithms. To this end, we treat the component as an evolvable object within the AAD process and refer to it as an Executable Algorithmic Primitive (EAP). \figurename\ \ref{fig:motivation_b} illustrates EAP. At the code level, EAP is an implemented callable function. During the AAD process, a promising EAP is extracted from an algorithm generated in an early iteration, stored in a persistent EAP set, and subsequently invoked by some candidates generated in later iterations.
EAP enters the set with an undetermined utility value. It updates the utility values by continuously competing with the existing EAPs. EAPs with higher utility values are more likely to be included in the newly generated algorithm.

To achieve EAP-enhanced AAD, we propose Primitive-Aware Code Evolution (PACE), which comprises two key mechanisms.
The first is a set of primitive-aware variation operators. On the one hand, the operators restrict LLMs to permuting EAPs within an algorithm, combining EAPs across algorithms, inserting EAPs into an algorithm, and replacing one EAP with another. On the other hand, any algorithmic component not explicitly represented as an EAP can be freely edited by LLMs.
The second is a selection rule that decides which EAP is exposed to the next variation step. PACE treats each EAP transfer as a single observation, recording whether the resulting offspring algorithm improves based on its parents. An EAP is then selected using Thompson sampling based on these observations. Consequently, the probability of selecting an EAP is dynamically adjusted as the AAD proceeds; EAPs that frequently fail are exposed less often.
Both mechanisms rely only on the algorithm evaluation, so no auxiliary validation set and no additional evaluation budget are required.

Our contributions are highlighted as follows:
\begin{itemize}

    \item We introduce EAPs to enable the explicit transfer of algorithmic local components.
    The EAP is an implemented function and has a stable identity across host algorithms. During the AAD process, an EAP is transferred across algorithms to accumulate evidence for updating its utility.

    \item We propose an EAP-characterized AAD method called PACE. Offspring generated by primitive-aware variation operators are associated with specific EAPs. PACE assigns credit to each EAP by comparing the corresponding offspring algorithms with their parents. Using Thompson sampling, PACE selectively leverages effective EAPs to accelerate the discovery of high-performing algorithms.
    \item Experiments on four tasks spanning continuous control and combinatorial optimization show that PACE finds better algorithms than existing LLM-based AAD methods under the same evaluation budget.
\end{itemize}

\section{Related Work}

\subsection{LLM-Based Automated Algorithm Design}
Since AEL and EoH introduced evaluator-guided program evolution using LLMs \cite{liu2023ael,liu2024eoh,liu2024survey}, research in this field has expanded rapidly in two main directions.
On the methodological side, recent algorithms improve the context carried across search iterations, incorporating written reflections \cite{ye2024reevo}, search trees \cite{zheng2025mctsahd}, or diverse populations \cite{dat2025hsevo}.
On the application side, this evolutionary loop has spread from heuristic design to math discovery \cite{romeraparedes2024funsearch}, system optimization \cite{novikov2025alphaevolve}, symbolic equation discovery \cite{shojaee2024llmsr}, and robot control, either indirectly by evolving reward functions \cite{ma2023eureka} or directly from execution feedback \cite{hu2026mles}.

Despite these advances in methods and applications, existing frameworks stick to full-program evolution.
Search memory stays inside external prompts, trajectories, or candidate pools, treating each generated algorithm as a single unit.
Overcoming this bottleneck requires storing and transferring smaller sub-program components across search runs.

\subsection{Reusable Structure in Program Search}
Reusing structural sub-components has a long history, including early genetic programming \cite{koza1992genetic,koza1994genetic}, library learning in DreamCoder \cite{ellis2021dreamcoder}, and modern LLM skill libraries \cite{wang2023voyager,stengeleskin2024regal,wang2023lego,grand2024lilo,wang2024trove}.
However, skill libraries usually accept components based on binary pass or fail tests.
This makes them unsuitable for algorithm design, where code quality is continuously scored and depends heavily on the task context.

Within algorithm design, recent studies explore structural reuse to move past single-candidate evolution.
Methods such as G-LNS \cite{zhao2026glns}, EoH-S \cite{liu2025eohs}, EvoLattice \cite{yuksel2025evolattice}, and BEAM \cite{xiang2026beam} save components using jointly evolving operators, fixed graph templates, or two-tier memory systems.
However, these methods require predefined candidate roles or heavy nested search loops.
In contrast, PACE establishes Extracted Algorithm Primitives (EAPs) as independent evolutionary units, enabling flexible reuse across unrelated program lineages without extra search overhead.

While transferring EAPs across host programs enables component reuse, it introduces an evaluation challenge.
Because an EAP executes inside a specific host program, the quality of that host can easily mask the true performance of the primitive itself.

\subsection{Bandit Credit Assignment}
Separating the contribution of a local function from its host program is a classic credit-assignment problem.
Evolutionary optimization often solves credit assignment using bandit feedback, such as adaptive operator selection \cite{fialho2010analyzing} and Thompson sampling \cite{russo2018tutorial}.
In LLM-based algorithm design, bandit strategies have been applied to whole programs, helping rank solution candidates in QUBE \cite{chen2024qube} or balance algorithm types in CDEoH \cite{wang2026cdeoh}.

However, existing bandit setups define arms over full programs, fixed variation operators, or set template slots.
PACE redefines the arm set over an open pool of EAPs that grows during search.
To isolate an EAP's causal impact from host program noise, PACE evaluates arms using parent-relative rewards, measuring performance gains over the fixed parent baseline upon transfer.

\section{Executable Algorithmic Primitive (EAP)}

\begin{figure*}[t]
    \centering
    \includegraphics[width=0.98\textwidth]{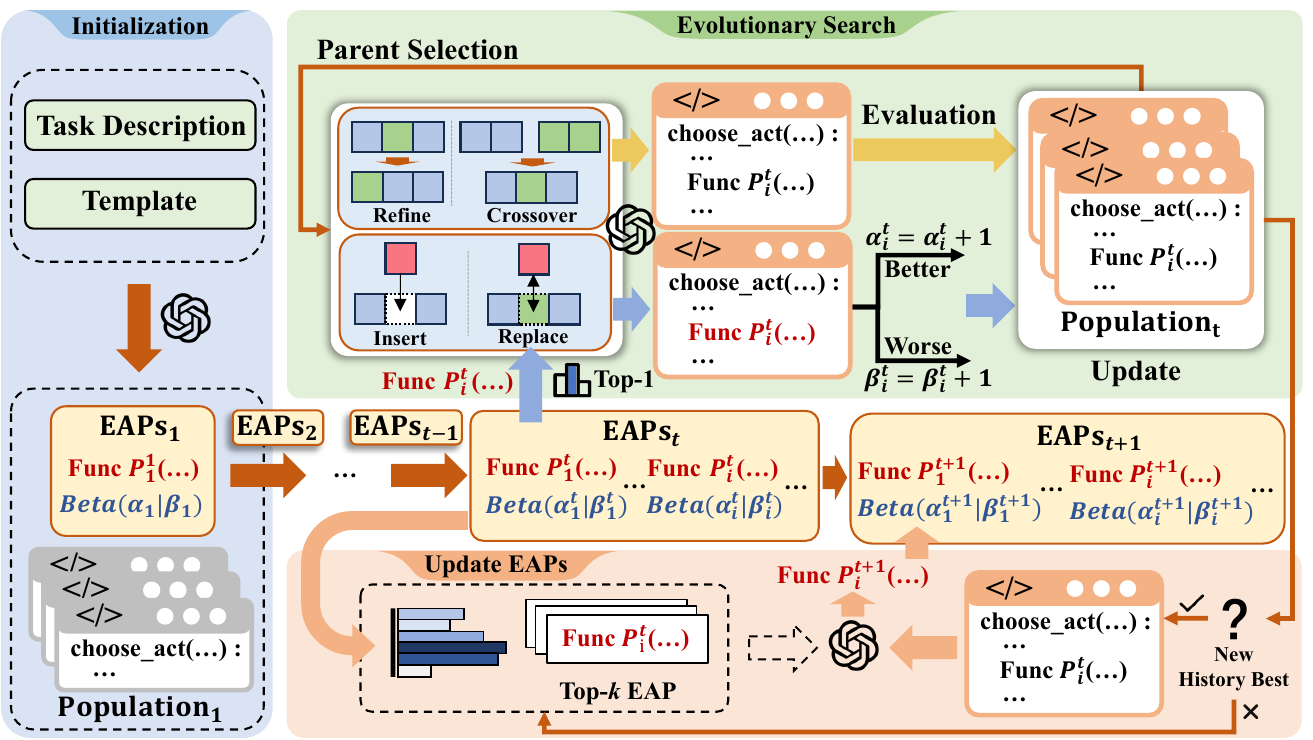}
    \caption{Overview of PACE. Complete algorithms follow an evolutionary loop of parent selection, generation, evaluation, and population update. Meanwhile, EAPs are discovered from the task or evaluated algorithms, retained independently, and selected for transfer. Operators \textit{Insert} and \textit{Replace} introduce one focus EAP and convert parent-relative performance into transfer evidence. Operators \textit{Refine} and \textit{Crossover} reuse existing EAP structure without assigning credit to a single EAP.}
    \label{fig:framework}
\end{figure*}

\subsection{From Program Primitives to EAPs}
The view that a program can be composed from reusable units has a long history in program synthesis and evolutionary computation. In the standard formulation of genetic programming, programs are constructed from a primitive set consisting of functions \cite{koza1992genetic}. These primitives define the elementary operations from which an evolutionary process builds a program. Automatically defined functions further allow an evolved program to reuse a discovered subprogram \cite{koza1994genetic}. In both cases, the primitives are either specified before search or represented as part of the individual program being evolved.

PACE adopts this compositional view but assigns a different role to the component. An EAP is a callable function that is generated during search and retained independently of the complete algorithm in which it was created or first observed. Formally, an EAP $e$ is
\begin{equation}
    e=(\sigma_e,\phi_e),
\end{equation}
where $\sigma_e$ specifies its textual description, and $\phi_e$ is its executable function. The implementation $\phi_e$ remains fixed when the EAP is transferred between algorithms; modifying it defines a different EAP.

Let $\mathcal{E}_t$ denote the EAPs available at search step $t$. For a complete algorithm $A$, its EAP call set is
\begin{equation}
    \mathcal{C}(A;\mathcal{E}_t)
    =\{e\in\mathcal{E}_t\mid A\text{ invokes }e\}.
\end{equation}
The definition separates availability from use. An EAP can remain in $\mathcal{E}_t$ without being called by a algorithm from the current population, while an explicit function call establishes its participation in a particular complete algorithm.

\subsection{Primitive-Aware Automated Algorithm Design}

Consider an AAD task with a space of algorithms $\mathcal{A}$ and an evaluator $J:\mathcal{A}\rightarrow\mathbb{R}$, where a larger value denotes a better algorithm. Under a finite evaluation budget $B$, standard AAD seeks
\begin{equation}
    A^{\star}=\arg\max_{A\in\mathcal{A}}J(A),
    \qquad N_{J}\leq B,
\end{equation}
where $N_{J}$ denotes the number of evaluator calls. The evaluator assigns scores only to complete algorithms. This objective remains unchanged in primitive-aware AAD.

PACE extends the state used to search for $A^{\star}$. At step $t$, its state is
\begin{equation}
    S_t=(\mathcal{P}_t,\mathcal{E}_t,\mathcal{H}_t),
\end{equation}
where $\mathcal{P}_t$ is a population of complete algorithms, $\mathcal{E}_t$ is the persistent EAP set, and $\mathcal{H}_t$ records evidence obtained when EAPs are transferred between algorithms. Complete algorithms remain the objects evaluated by $J$. EAPs have no separate task objective. Instead, $\mathcal{H}_t$ affects which EAPs are exposed to later variation. PACE therefore changes the search state and the variation process. 

The two search objects have distinct lifetimes. Population selection may remove $A$ from $\mathcal{P}_t$, but this removal does not delete an EAP previously obtained from $A$. For every admitted EAP,
\begin{equation}
    e\in\mathcal{E}_t \ \Longrightarrow\ e\in\mathcal{E}_{t+1}.
\end{equation}
This persistence makes the local function available to algorithms that are generated after its source algorithm has disappeared.

\section{Primitive-Aware Code Evolution (PACE)} \label{sec:method}

\subsection{Overview}

\figurename\ \ref{fig:framework} presents the interaction between complete-algorithm evolution and EAP adaptation. PACE first initializes an algorithm population and an EAP from task description and algorithm template. During evolution, it repeatedly selects parent algorithms, assigns EAPs to one of four primitive-aware operators, and asks the LLM to produce a complete child algorithm. The task evaluator scores the child and updates the population in the same manner as population-based AAD.

The EAP process supplies a second source of search memory. New EAP is introduced either by extending from tok-k EAP or extracting from the new evaluated history best algorithm, adaptive selection chooses which EAP to transfer, and the primitive-aware operators control how that transfer occurs. When an operator introduces one focus EAP, the child is compared with its fixed parent and the result updates the EAP's transfer history. This evidence subsequently changes its probability of selection. The two processes are coupled through evaluated complete algorithms, so PACE obtains EAP feedback without a separate evaluation objective or an additional validation set.

\subsection{Adaptive EAP Selection} \label{sec:credit}
An EAP can exhibit varying utility depending on the parent algorithm it is injected into. Therefore, rather than assigning a static, context-independent score to an EAP, PACE evaluates its empirical transferability through focused transfer trials. Let a trial consist of a parent algorithm $A$, a generated child $A'$, and the explicitly injected EAP $e$. The trial yields a binary reward:
\begin{equation}
    r_e = \mathbb{I}\!\left[J(A') > J(A)\right].
\end{equation}
We model $r_e$ as a realization of a Bernoulli random variable parameterized by an unknown success rate $\rho_e$. Formally, $\rho_e$ is defined as the conditional probability of achieving a strict performance improvement given the explicit injection of $e$:
\begin{equation}
    \rho_e = \mathbb{P}\big(J(A') > J(A) \mid e \in \mathcal{C}(A')\big).
\end{equation}

To continuously estimate $\rho_e$ in an online manner, PACE maintains a Beta posterior for each EAP. Initialized with an uninformative prior $\mathrm{Beta}(1,1)$, the posterior parameters $(\alpha_e, \beta_e)$ are sequentially updated after each valid trial:
\begin{equation}
    (\alpha_e, \beta_e) \leftarrow (\alpha_e + r_e, \, \beta_e + 1 - r_e).
\end{equation}
Specifically, a strictly improving child increments $\alpha_e$. A degradation, score tie, or runtime execution error increments $\beta_e$. Notably, if the LLM fails to structurally integrate $e$ into $A'$, the trial is deemed invalid, and the posterior remains unchanged to prevent biased penalty.

To seamlessly balance the exploitation of highly transferable EAPs with the exploration of uncertain ones, PACE formulates this selection process as a Multi-Armed Bandit (MAB) \cite{slivkins2019mab} problem and resolves it using Thompson Sampling (TS) \cite{russo2018tutorial}. When an operator requires injecting an EAP into parent $A$, let $\mathcal{G}_t(A) \subseteq \mathcal{E}_t \setminus \mathcal{C}(A)$ denote the eligible candidates. For each $e \in \mathcal{G}_t(A)$, PACE independently draws a belief sample $\theta_e$ from its posterior:
\begin{equation}
    \theta_e \sim \mathrm{Beta}(\alpha_e, \beta_e),
\end{equation}
and dynamically selects the EAP with the maximum $\theta_e$. To prevent the premature starvation of newly extracted EAPs caused by a lack of observations, they undergo a brief, forced warm-up exposure before being subjected to standard Thompson selection.

\subsection{Primitive-Aware Operators} \label{subsec:op}

To effectively explore the algorithmic space, we introduces four primitive-aware variation operators, invoked with equal probability during the search. While each operator prompts the LLM to generate a complete child algorithm, they enforce distinct structural constraints on the inheritance and modification of EAP calls.

Let $\mathcal{V} \subseteq \mathcal{E}_t$ denote the designated subset of candidate EAPs exposed to the LLM in the current operational prompt. We define $\mathcal{C}_{\mathcal{V}}(A) = \mathcal{C}(A) \cap \mathcal{V}$ to represent the active primitives invoked by program $A$ strictly within this exposed context. The four operator contracts below mandate the structural composition of the child program $A'$ over $\mathcal{C}_{\mathcal{V}}(\cdot)$.

\paragraph{P1: Primitive Insertion.}
Given a parent $A$ with an available EAP slot, Thompson sampling selects a focus EAP $e^{+}\notin\mathcal{C}_{\mathcal{V}}(A)$. The LLM must integrate $e^+$ while preserving all previously exposed parent EAPs:
\begin{equation}
    \mathcal{C}_{\mathcal{V}}(A')=
    \mathcal{C}_{\mathcal{V}}(A)\cup\{e^{+}\}, \quad |\mathcal{C}_{\mathcal{V}}(A)| < K.
\end{equation}
The algorithm may be reorganized to integrate the new function. Since $e^{+}$ is the only newly introduced EAP, the parent-child comparison produces one focused transfer observation for $e^{+}$.

\paragraph{P2: Primitive Replacement.}
PACE selects one called EAP $e^{-}$ as the removal target and selects an absent EAP $e^{+}$ by Thompson sampling. All other parent EAPs are preserved:
\begin{equation}
    \mathcal{C}_{\mathcal{V}}(A')=
    \bigl(\mathcal{C}_{\mathcal{V}}(A)\setminus\{e^{-}\}\bigr)
    \cup\{e^{+}\}.
\end{equation}
The removal target is the called EAP with the lowest posterior mean. The focused credit observation from this trial is strictly assigned to the injected EAP $e^+$, evaluating its capability to substitute the incumbent local logic.

\paragraph{P3: Primitive-Preserving Refinement.}
This operator optimizes how the current EAP composition is utilized without altering its membership:
\begin{equation}
    \mathcal{C}_{\mathcal{V}}(A') = \mathcal{C}_{\mathcal{V}}(A).
\end{equation}
The LLM is prompted to refine order, parameters, or interactions around the fixed EAP calls. This cleanly separates the discovery of optimal primitive combinations from the structural adaptation of the macro-algorithm. Since no new EAP is introduced, this operation does not trigger a posterior update.

\paragraph{P4: Primitive-Aware Crossover.}
PACE selects two parent algorithms $A_1$ and $A_2$. This operator exposes the complete union of active primitives from both parents to the LLM. To guarantee a genuine cross-program composition, the generated child $A'$ must freely recombine a subset of this joint pool under the maximum capacity $K$, while strictly inheriting at least one primitive from each parent:
\begin{equation}
    \begin{aligned}
        & \mathcal{C}_{\mathcal{V}}(A') \subseteq \mathcal{C}_{\mathcal{V}}(A_1) \cup \mathcal{C}_{\mathcal{V}}(A_2), \quad |\mathcal{C}_{\mathcal{V}}(A')| \le K, \\
        & \mathcal{C}_{\mathcal{V}}(A') \cap \mathcal{C}_{\mathcal{V}}(A_1) \neq \emptyset, \quad \mathcal{C}_{\mathcal{V}}(A') \cap \mathcal{C}_{\mathcal{V}}(A_2) \neq \emptyset.
    \end{aligned}
\end{equation}
The LLM is granted the autonomy to reorganize the overarching algorithmic structure to support these interacting primitives. However, P4 also does not trigger a posterior update for any EAP, as the simultaneous unconstrained recombination of multiple primitives fundamentally confounds credit assignment.

The call-set relations defined above operate as strict structural constraints rather than optional prompt. PACE programmatically verifies the required, preserved, and removed primitive calls via the AST verifier before validating a candidate as a realized operator outcome. Candidates failing to satisfy their assigned structural contracts are immediately discarded from the Thompson sampling update, ensuring that unverified LLM hallucinations do not corrupt the empirical transfer evidence.

\subsection{EAP Discovery} \label{subsec:eap-discovery}

PACE discovers and populates new EAPs through two mutually exclusive mechanisms executed at the end of each generation:

\begin{itemize}
    \item \textbf{EAP Generation.} Rather than exploring the functional space blindly, this mechanism leverages proven historical discoveries to guide the creation of novel logic. Conditioned on the task specification and the implementations of the top-$k$ performing EAPs, the LLM is explicitly prompted to synthesize a functionally distinct EAP. By referencing these high-quality primitives, the generation process is forced to extrapolate beyond existing capabilities and avoid functional redundancy. 

    \item \textbf{EAP Extraction.} This mechanism refactors local logic from a newly evaluated historical best into a persistent EAP, strictly without altering or reevaluating the source program. Because an incumbent best will likely be superseded as the search progresses, this strategy acts as a preservation mechanism. As illustrated in \figurename\ \ref{fig:motivation}(b), it rescues valuable local logic by isolating it as an independent entity for future cross-program transfer, ensuring it survives even after its original algorithm is eventually discarded. While extraction leverages high-quality algorithms, it does not guarantee that the extracted component solely caused the performance gain. Subsequent focused trials empirically evaluate its true transferability.
\end{itemize}

To maintain the compactness of the EAP library and prevent exceeding the evaluation budget, PACE enforces strict governance rules: at most one EAP is proposed per generation, and EAP generation and extraction are strictly mutually exclusive, with extraction holding higher priority.

\section{Experiments}
\label{sec:experiments}

\subsection{Experimental Setup}
\paragraph{Benchmarks.}
We evaluate PACE across two distinct types of tasks.
\begin{itemize}
  \item \textbf{Continuous Control.} Following MLES \cite{hu2026mles}, solving these OpenAI Gym environments \cite{brockman2016openai} entails evolving programmatic algorithms to replace neural policies. We evaluate Racing Car for visual inputs and extend this paradigm to Bipedal Walker for state inputs to test whether EAPs generalize across different control tasks.
  \item \textbf{Combinatorial Optimization.} We evaluate on the Traveling Salesperson Problem (TSP) and TSP guided by Ant Colony Optimization (TSP-ACO). These tasks require evolving algorithms to improve final routing quality across complex search spaces.
\end{itemize}

\begin{figure*}[t]
    \centering
    \begin{subfigure}[b]{0.32\linewidth}
        \centering
        \includegraphics[width=\linewidth]{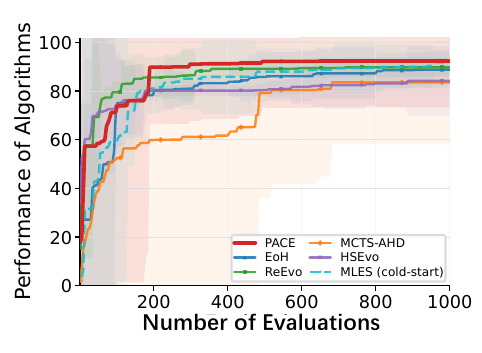}
        \caption{Racing Car}
        \label{fig:conv-carracing}
    \end{subfigure}
    \hfill
    \begin{subfigure}[b]{0.32\linewidth}
        \centering
        \includegraphics[width=\linewidth]{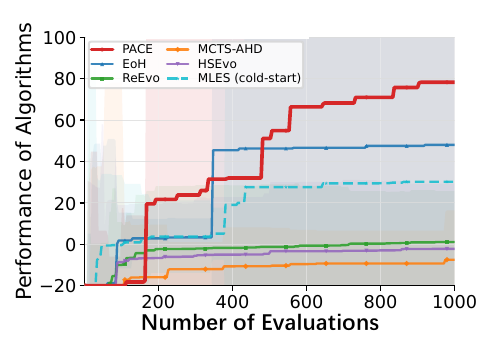}
        \caption{Bipedal Walker}
        \label{fig:conv-bipedalwalker}
    \end{subfigure}
    \hfill
    \begin{subfigure}[b]{0.32\linewidth}
        \centering
        \includegraphics[width=\linewidth]{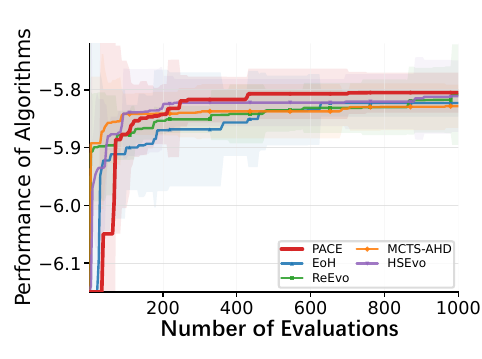}
        \caption{TSP-ACO}
        \label{fig:conv-tsp-aco}
    \end{subfigure}
    \caption{Convergence curves on three diverse tasks.}
    \label{fig:convergence}
\end{figure*}

\paragraph{Baselines}

We compare PACE against three types of baselines.
\begin{itemize}
  \item \textbf{LLM-based AAD Methods.} This type includes EoH \cite{liu2024eoh} for standard evolution, ReEvo \cite{ye2024reevo} for reflective evolution, MCTS-AHD \cite{zheng2025mctsahd} for tree search, and HSEvo \cite{dat2025hsevo} for hybrid diversity. We also include MLES \cite{hu2026mles} for multimodal control evolution, strictly using its cold-start mode because the seeded mode introduces unfair prior knowledge through external programs.
  \item \textbf{Neural Methods.} This category includes PPO \cite{schulman2017ppo} for control tasks, alongside DeepACO \cite{ye2023deepaco} for combinatorial optimization. We adopt PPO settings directly from MLES for fair comparisons.
  \item \textbf{Classical Heuristics.} We include traditional ACO \cite{dorigo2006ant} as a non-learning baseline for TSP-ACO.
\end{itemize}

For experiment in these section, we set the only parameter $k=3$ in PACE. All AAD methods use GPT-4o-mini for algorithm generation. Each method conducts three independent runs of 1000 evaluations. We report the mean and standard deviation of the best training scores. We evaluate the best overall training program on the test set. All experiments ran on Intel Xeon Gold 6348 CPUs. 

\begin{table}[t]
\centering
\small
\setlength{\tabcolsep}{2.5pt}
\caption{Training and testing performance on the control tasks. Bold indicates the best result among AAD methods.}
\begin{tabular}{
  l 
  S[table-format=2.2]@{\,$\pm$\,}S[table-format=2.2] 
  S[table-format=2.2] 
  S[table-format=-1.2]@{\,$\pm$\,}S[table-format=2.2] 
  S[table-format=2.2]
}
\toprule
& \multicolumn{3}{c}{Racing Car $\uparrow$}
& \multicolumn{3}{c}{Bipedal Walker $\uparrow$} \\
\cmidrule(lr){2-4}\cmidrule(lr){5-7}
Method & \multicolumn{2}{c}{Training} & \multicolumn{1}{c}{Test} & \multicolumn{2}{c}{Training} & \multicolumn{1}{c}{Test} \\
\midrule
PPO & 94.21 & 1.51 & 85.69 & 314.70 & 6.09 & 322.38 \\
\midrule
EoH & 88.81 & 3.00 & 64.64 & 48.06 & 73.53 & -2.61 \\
ReEvo & 89.77 & 1.60 & 66.20 & 1.09 & 9.91 & 2.03 \\
MCTS-AHD & 83.56 & 10.24 & 73.52 & -7.45 & 9.58 & 2.47 \\
HSEvo & 84.17 & 5.57 & 84.17 & -2.23 & 12.72 & 8.97 \\
MLES (cold) & 91.90 & 2.17 & 83.59 & 30.25 & 39.05 & 15.42 \\
\rowcolor{gray!15}
PACE & \textbf{92.40} & \textbf{7.69} & \textbf{98.90} & \textbf{67.06} & \textbf{72.63} & \textbf{133.71} \\
\bottomrule
\end{tabular}
\label{tab:control-results}
\end{table}


\begin{table}[t]
\centering
\small
\setlength{\tabcolsep}{5pt}
\caption{Testing result on TSP-Construct. The underlined size is the in-domain scale. The remaining columns evaluate that same program on out-domain scale. Bold indicates the best result at each scale.}
\begin{tabular}{lcccc}
\toprule
& \multicolumn{4}{c}{TSP-Construct$\downarrow$} \\
\cmidrule(lr){2-5}
Method & $\underline{n=50}$ & $n=200$ & $n=500$ & $n=1000$ \\
\midrule
Greedy Construct & 6.938 & 13.362 & 20.794 & 28.917\\
\midrule
EoH & 6.279 & 12.255 & 19.035 & 26.862 \\
ReEvo & 6.328 & 12.198 & 19.161 & 27.070 \\
MCTS-AHD & 6.199 & 12.136 & 18.997 & 26.876 \\
HSEvo & 6.218 & 12.388 & 19.506 & 27.326 \\
\rowcolor{gray!15}
PACE & \textbf{6.013} & \textbf{12.064} & \textbf{18.825} & \textbf{26.596} \\
\bottomrule
\end{tabular}
\label{tab:tsp-results}
\end{table}

\subsection{Comparison Results}

\paragraph{Continuous Control Task.}

We evaluate PACE on continuous control tasks, including Racing Car and Bipedal Walker.
Table \ref{tab:control-results} reports training and testing performance, while \figurename\ \ref{fig:conv-carracing} and \figurename\ \ref{fig:conv-bipedalwalker} illustrate search convergence curves across 1,000 evaluations.

To contextualize these results, we distinguish between black-box neural RL (PPO) and programmatic code generation (AAD methods including PACE).

On Racing Car, Table \ref{tab:control-results} shows that PACE achieves a training score of 92.40 and a test score of 98.90.
Remarkably, PACE outperforms not only all AAD baselines like HSEvo at 84.17, but also neural PPO at 85.688 in zero-shot test generalization.
As shown in \figurename\ \ref{fig:conv-carracing}, PACE converges rapidly within the first 200 evaluations and maintains this lead throughout the search.

On Bipedal Walker, neural PPO achieves a test score of 322.383, benefiting from the expressive capacity of dense neural networks to fit complex motor control.
However, within the scope of programmatic algorithm design, existing AAD methods struggle severely.
Baselines like ReEvo, HSEvo, and MCTS-AHD flatline near zero or negative test scores due to premature convergence.
EoH achieves 48.06 in training but overfits to -2.61 on testing.
In contrast, PACE reaches a training score of 67.06 and an impressive test score of 133.71, outperforming the specialized control baseline MLES cold-start at 15.42 by a wide margin.

The low variance of baselines merely reflects their consistent failure to find viable controllers. Exploring this multi-modal space inherently increases variance, yet PACE composes functional control primitives where full-program search fails, setting a new state-of-the-art for programmatic policy search.

\paragraph{Combinatorial Optimization Task.}
We first evaluate PACE on the TSP-ACO benchmark. Following the experimental setup in MCTS-AHD \cite{zheng2025mctsahd}, all algorithms evolve on the in-domain scale of $n=50$, and the best evolved programs are evaluated zero-shot on out-domain scales of $n=200$, $n=500$, and $n=1000$ with 64 instances. \figurename\ \ref{fig:conv-tsp-aco} shows the search convergence curves, where PACE consistently achieves superior optimization performance.

Table \ref{tab:tspaco-results} presents the comparative results.
On the in-domain scale of $n=50$, all LLM-based methods outperform classical ACO and DeepACO. ReEvo obtains a slightly lower cost of 5.774, while PACE achieves a competitive cost of 5.795. However, significant differences arise when applying the evolved programs to larger problem scales. PACE consistently achieves the best performance across all out-domain scales, reaching 11.645 at $n=200$, 19.485 at $n=500$, and 28.130 at $n=1000$. In contrast, baseline methods exhibit performance degradation during scale transfer. Most notably, MCTS-AHD performs competitively at $n=50$ but suffers massive degradation as problem size increases, yielding a cost of 50.291 at $n=1000$, which is worse than classical non-learning ACO.

We next evaluate PACE on the TSP-Construct benchmark. Following the same zero-shot evaluation protocol, algorithms are evolved on the in-domain scale of $n=50$ and evaluated on larger out-domain scales of $n=200$, $n=500$, and $n=1000$ with 64 instances. Table \ref{tab:tsp-results} presents the comparative results across all scales. All AAD methods outperform Greedy Construction. PACE achieves the lowest tour cost across all problem sizes on TSP-Construct. On the in-domain scale of $n=50$, PACE achieves a cost of 6.013, outperforming the second-best baseline MCTS-AHD at 6.199. This advantage persists during zero-shot scale transfer, where PACE consistently leads at $n=200$ with 12.064, $n=500$ with 18.825, and $n=1000$ with 26.596.

\subsection{Ablation Studies}
\paragraph{Ablation on Parameters and Components.}
We first remove Thompson Sampling and select primitives randomly instead. Table \ref{tab:mechanism-ablation} shows a severe performance drop across both tasks. Preserving primitives alone is clearly insufficient. The search mechanism must intelligently decide which primitive to transfer.

\begin{table}[t]
\centering
\setlength{\tabcolsep}{5pt}
\caption{Testing result on TSP-ACO. The underlined size is the in-domain scale. The remaining columns evaluate that same program on out-domain scale. Bold indicates the best result at each scale.}
\begin{tabular}{lcccc}
\toprule
& \multicolumn{4}{c}{TSP-ACO$\downarrow$} \\
\cmidrule(lr){2-5}
Method & $\underline{n=50}$ & $n=200$ & $n=500$ & $n=1000$ \\
\midrule
ACO & 6.664 & 15.479 & 26.926 & 41.111\\
DeepACO & 5.845 & 12.188 & 21.573 & 37.308\\
\midrule
EoH & $5.812$ & $12.382$ & $21.008$ & $30.733$ \\
ReEvo & $\mathbf{5.774}$ & $11.725$ & $19.644$ & $28.305$ \\
MCTS-AHD & $5.811$ & $14.769$ & $23.936$ & $50.291$ \\
HSEvo & $5.783$ & $11.888$ & $20.167$ & $30.081$ \\
\rowcolor{gray!15}
PACE & $5.795$ & $\mathbf{11.645}$ & $\mathbf{19.485}$ & $\mathbf{28.130}$ \\
\bottomrule
\end{tabular}
\label{tab:tspaco-results}
\end{table}

The primitive operators are equally essential. Table \ref{tab:mechanism-ablation} indicates that removing any operator from P1 to P4 degrades algorithmic performance. Operator P3 is particularly critical. Removing P3 causes the score on Racing Car to crash to 79.413, and degrades the TSP routing score to 5.832. This proves these operators effectively manage primitive reuse.

We also evaluate the extraction and generation modules in Table \ref{tab:mechanism-ablation}. Removing primitive extraction harms the final quality across both tasks. Removing primitive generation causes a noticeable performance drop on Racing Car. The routing score on TSP shows a very small fluctuation (a difference of 0.009). This minor difference easily falls within the natural noise of automated algorithm design. The generation module remains highly effective on at least one complex task. This confirms the overall necessity of both modules.

We finally measure the sensitivity of the only parameter $k$ in PACE. This parameter bounds the maximum number of primitives per algorithm. The last two rows of Table \ref{tab:mechanism-ablation} show the results. The default setting ($k=3$) yields the best overall balance. A smaller value ($k=1$) restricts algorithmic expressiveness. Conversely, a larger value ($k=5$) introduces excessive context noise and degrades performance.

\begin{table}[t]
\centering
\setlength{\tabcolsep}{6pt}
\caption{Ablation on Racing Car and TSP-ACO.}
\begin{tabular}{l r@{\,$\pm$\,}l r@{\,$\pm$\,}l}
\toprule
Method & 
\multicolumn{2}{c}{Racing Car$\uparrow$} &
\multicolumn{2}{c}{TSP-ACO$\downarrow$} \\
\midrule
PACE & 92.395 & 7.690 & 5.805 & 0.014 \\
\midrule
w/o TS & 85.058 & 9.180 & 5.815 & 0.025\\
\midrule
w/o P1  & 90.974 & 1.689 & 5.818 & 0.012 \\
w/o P2  & 91.207 & 2.305 & 5.824 & 0.006 \\
w/o P3  & 79.413 & 9.622 & 5.832 & 0.025 \\
w/o P4  & 91.251 & 5.657 & 5.824 & 0.009 \\
\midrule
w/o EAP Generation & 89.555 & 2.401 & 5.796 & 0.017 \\
w/o EAP Extract & 90.405 & 2.199 & 5.807 & 0.014 \\
\midrule
$k = 1$ & 91.801 & 4.250 & 5.821 & 0.001 \\
$k = 5$ & 89.350 & 6.105 & 5.834 & 0.019 \\
\bottomrule
\end{tabular}
\label{tab:mechanism-ablation}
\end{table}

\paragraph{Ablation on LLMs.}
We evaluate the robustness of PACE across different LLM backbones.
Because PACE decouples EAP discovery from host algorithm generation, distinct LLMs can be assigned to each search phase.
As shown in Table \ref{tab:model-allocation}, upgrading the EAP discovery model while keeping host algorithm generation on GPT-4o-mini produces substantial performance improvements.
Crucially, EAP discovery accounts for on average only 1.6\% of total token consumption during search.
This result demonstrates that enhancing model capability solely within the primitive discovery phase yields disproportionate gains for the entire search, validating the advantage of representing EAPs independently.
We also evaluate unified setups where a single LLM handles both phases.
PACE exhibits strong robustness across diverse backbones, reaching an average score of 99.005 on Racing Car, approaching the maximum score of 100 when powered by Gemini-3.1-flash-lite.

\begin{table}[t]
\centering
\small
\setlength{\tabcolsep}{5pt}
\caption{Result of ensemble model on Racing Car.}
\begin{tabular}{
  l 
  l 
  S[table-format=2.3]@{\,$\pm$\,}S[table-format=1.3]
}
\toprule
Algorithm & EAP & \multicolumn{2}{c}{Racing Car$\uparrow$} \\
\midrule
GPT-4o-mini & DeepSeek-v4-flash &  94.597 &  3.865 \\
GPT-4o-mini & Gemini-3.1-flash-lite & 93.448 &  2.508 \\
\midrule
GPT-4o-mini & GPT-4o-mini & 92.395 & 7.690 \\
DeepSeek-v4-flash & DeepSeek-v4-flash & 94.539 & 1.543 \\
Gemini-3.1-flash-lite & Gemini-3.1-flash-lite & 99.005 & 0.580 \\
\bottomrule
\end{tabular}
\label{tab:model-allocation}
\end{table}

\section{Conclusion}

In this paper, we introduce EAPs to decouple reusable local logic from complete algorithms. Building on this representation, we present PACE. PACE maintains a dynamic set of EAPs and applies specialized operators to structurally guarantee their cross-algorithm transfer. The framework leverages Thompson Sampling to guide EAP selection based on relative performance improvements. Experiments across continuous control and combinatorial optimization show that PACE achieves superior performance compared to existing baseline methods. The results confirm that preserving and transferring valuable local primitives can enhance final algorithmic performance. PACE also generalizes well across expanding problem scales and dynamic environments. Ultimately, this work shifts automated algorithm design from full-program search toward modular composition.

\paragraph{Limitation and future work.}
PACE assumes that primitives can be evaluated and selected independently. In algorithms with extremely complex constraints, strong coupling may exist between different primitives. Evaluating primitives separately might not fully capture their joint interactions. Future work will investigate primitive coupling metrics to explicitly model dependencies between EAPs during search.

\clearpage

\bibliography{aaai2027}

@article{romeraparedes2024funsearch,
  title={Mathematical discoveries from program search with large language models},
  author={Romera-Paredes, Bernardino and Barekatain, Mohammadamin and Novikov, Alexander and Balog, Matej and Kumar, M Pawan and Dupont, Emilien and Ruiz, Francisco JR and Ellenberg, Jordan S and Wang, Pengming and Fawzi, Omar and others},
  journal={Nature},
  volume={625},
  number={7995},
  pages={468--475},
  year={2024},
  publisher={Nature Publishing Group UK London}
}

@inproceedings{liu2024eoh,
  title={Evolution of heuristics: towards efficient automatic algorithm design using large language model},
  author={Liu, Fei and Tong, Xialiang and Yuan, Mingxuan and Lin, Xi and Luo, Fu and Wang, Zhenkun and Lu, Zhichao and Zhang, Qingfu},
  booktitle={Proceedings of the 41st International Conference on Machine Learning},
  pages={32201--32223},
  year={2024}
}

@article{liu2023ael,
  title={Algorithm evolution using large language model},
  author={Liu, Fei and Tong, Xialiang and Yuan, Mingxuan and Zhang, Qingfu},
  journal={arXiv preprint arXiv:2311.15249},
  year={2023}
}

@inproceedings{ye2024reevo,
  title={ReEvo: large language models as hyper-heuristics with reflective evolution},
  author={Ye, Haoran and Wang, Jiarui and Cao, Zhiguang and Berto, Federico and Hua, Chuanbo and Kim, Haeyeon and Park, Jinkyoo and Song, Guojie},
  booktitle={Proceedings of the 38th International Conference on Neural Information Processing Systems},
  pages={43571--43608},
  year={2024}
}

@inproceedings{zheng2025mctsahd,
  title={Monte Carlo Tree Search for Comprehensive Exploration in LLM-Based Automatic Heuristic Design},
  author={Zheng, Zhi and Xie, Zhuoliang and Wang, Zhenkun and Hooi, Bryan},
  year={2025},
  booktitle={Forty-second International Conference on Machine Learning}
}

@article{liu2024survey,
  title={A systematic survey on large language models for algorithm design},
  author={Liu, Fei and Yao, Yiming and Guo, Ping and Yang, Zhiyuan and Lin, Xi and Zhao, Zhe and Tong, Xialiang and Mao, Kun and Lu, Zhichao and Wang, Zhenkun and others},
  journal={ACM Computing Surveys},
  volume={58},
  number={8},
  pages={1--32},
  year={2026},
  publisher={ACM New York, NY}
}

@inproceedings{liu2025eohs,
  title={Eoh-s: Evolution of heuristic set using llms for automated heuristic design},
  author={Liu, Fei and Liu, Yilu and Zhang, Qingfu and Xialiang, Tong and Yuan, Mingxuan},
  booktitle={Proceedings of the AAAI Conference on Artificial Intelligence},
  number={43},
  pages={37090--37098},
  year={2026}
}

@article{novikov2025alphaevolve,
  title={Alphaevolve: A coding agent for scientific and algorithmic discovery},
  author={Novikov, Alexander and V{\~u}, Ng{\^a}n and Eisenberger, Marvin and Dupont, Emilien and Huang, Po-Sen and Wagner, Adam Zsolt and Shirobokov, Sergey and Kozlovskii, Borislav and Ruiz, Francisco JR and Mehrabian, Abbas and others},
  journal={arXiv preprint arXiv:2506.13131},
  year={2025}
}

@inproceedings{shojaee2024llmsr,
  title={Llm-sr: Scientific equation discovery via programming with large language models},
  author={Shojaee, Parshin and Meidani, Kazem and Gupta, Shashank and Barati Farimani, Amir and Reddy, Chandan},
  booktitle={International Conference on Learning Representations},
  volume={2025},
  pages={16054--16085},
  year={2025}
}

@inproceedings{ma2023eureka,
  title={Eureka: Human-level reward design via coding large language models},
  author={Ma, Yecheng Jason and Liang, William and Wang, Guanzhi and Huang, De-An and Bastani, Osbert and Jayaraman, Dinesh and Zhu, Yuke and Fan, Jim and others},
  booktitle={International conference on learning Representations},
  volume={2024},
  pages={26516--26560},
  year={2024}
}

@article{hu2026mles,
  title={Multimodal LLM-assisted Evolutionary Search for Programmatic Control Policies},
  author={Hu, Qinglong and Tong, Xialiang and Yuan, Mingxuan and Liu, Fei and Lu, Zhichao and Zhang, Qingfu},
  journal={arXiv preprint arXiv:2508.05433},
  year={2025}
}

@inproceedings{dat2025hsevo,
  title={Hsevo: Elevating automatic heuristic design with diversity-driven harmony search and genetic algorithm using llms},
  author={Dat, Pham Vu Tuan and Doan, Long and Binh, Huynh Thi Thanh},
  booktitle={Proceedings of the AAAI Conference on Artificial Intelligence},
  number={25},
  pages={26931--26938},
  year={2025}
}

@article{chen2024qube,
  title={QUBE: Enhancing Automatic Heuristic Design via Quality-Uncertainty Balanced Evolution},
  author={Chen, Zijie and Zhou, Zhanchao and Lu, Yu and Xu, Renjun and Pan, Lili and Lan, Zhenzhong},
  journal={arXiv preprint arXiv:2412.20694},
  year={2024}
}

@article{wang2026cdeoh,
  title={CDEoH: Category-Driven Automatic Algorithm Design With Large Language Models},
  author={Wang, Yu-Nian and Lyu, Shen-Huan and Chen, Ning and Xu, Jia-Le and Ye, Baoliu and Zhang, Qingfu},
  journal={arXiv preprint arXiv:2603.19284},
  year={2026}
}

@article{zhao2026glns,
  title={G-LNS: Generative large neighborhood search for LLM-based automatic heuristic design},
  author={Zhao, Baoyun and Wang, He and Zeng, Liang},
  journal={arXiv preprint arXiv:2602.08253},
  year={2026}
}

@article{russo2018tutorial,
  title={A tutorial on thompson sampling},
  author={Daniel, J Russo and Benjamin, Van Roy and Abbas, Kazerouni and Ian, Osband and Zheng, Wen},
  journal={Foundations and Trends{\textregistered} in Machine Learning},
  volume={11},
  number={1},
  pages={1--99},
  year={2018},
  publisher={Emerald Publishing Limited}
}

@book{koza1992genetic,
  title={Genetic programming: A paradigm for genetically breeding populations of computer programs to solve problems},
  author={Koza, John R},
  volume={34},
  year={1990},
  publisher={Stanford University, Department of Computer Science Stanford, CA}
}

@book{koza1994genetic,
  title={Genetic programming II: automatic discovery of reusable programs},
  author={Koza, John R},
  year={1994},
  publisher={MIT press}
}

@inproceedings{ellis2021dreamcoder,
  title={Dreamcoder: Bootstrapping inductive program synthesis with wake-sleep library learning},
  author={Ellis, Kevin and Wong, Catherine and Nye, Maxwell and Sabl{\'e}-Meyer, Mathias and Morales, Lucas and Hewitt, Luke and Cary, Luc and Solar-Lezama, Armando and Tenenbaum, Joshua B},
  booktitle={Proceedings of the 42nd acm sigplan international conference on programming language design and implementation},
  pages={835--850},
  year={2021}
}

@article{slivkins2019mab,
  title={Introduction to multi-armed bandits},
  author={Slivkins, Aleksandrs},
  journal={Foundations and Trends{\textregistered} in Machine Learning},
  volume={12},
  number={1-2},
  pages={1--286},
  year={2019},
  publisher={Emerald Publishing Limited}
}

@article{fialho2010analyzing,
  title={Analyzing bandit-based adaptive operator selection mechanisms},
  author={Fialho, {\'A}lvaro and Da Costa, Luis and Schoenauer, Marc and Sebag, Michele},
  journal={Annals of Mathematics and Artificial Intelligence},
  volume={60},
  number={1},
  pages={25--64},
  year={2010},
  publisher={Springer}
}

@article{yuksel2025evolattice,
  title={EvoLattice: Persistent Internal-Population Evolution through Multi-Alternative Quality-Diversity Graph Representations for LLM-Guided Program Discovery},
  author={Yuksel, Kamer Ali},
  journal={arXiv preprint arXiv:2512.13857},
  year={2025}
}

@article{xiang2026beam,
  title={BEAM: Bi-level Memory-adaptive Algorithmic Evolution for LLM-Powered Heuristic Design},
  author={Xiang, Chuyang and Wei, Yichen and Ma, Jiale and Wang, Handing and Yan, Junchi},
  journal={arXiv preprint arXiv:2604.12898},
  year={2026}
}

@article{wang2023lego,
  title={Lego-prover: Neural theorem proving with growing libraries},
  author={Wang, Haiming and Xin, Huajian and Zheng, Chuanyang and Li, Lin and Liu, Zhengying and Cao, Qingxing and Huang, Yinya and Xiong, Jing and Shi, Han and Xie, Enze and others},
  journal={arXiv preprint arXiv:2310.00656},
  year={2023}
}

@article{wang2024trove,
  title={Trove: Inducing verifiable and efficient toolboxes for solving programmatic tasks},
  author={Wang, Zhiruo and Fried, Daniel and Neubig, Graham},
  journal={arXiv preprint arXiv:2401.12869},
  year={2024}
}

@inproceedings{grand2024lilo,
  title={Lilo: Learning interpretable libraries by compressing and documenting code},
  author={Grand, Gabriel and Wong, Lio and Bowers, Maddy and Olausson, Theo X and Liu, Muxin and Tenenbaum, Joshua B and Andreas, Jacob},
  booktitle={International Conference on Learning Representations},
  volume={2024},
  pages={30399--30446},
  year={2024}
}

@article{stengeleskin2024regal,
  title={Regal: Refactoring programs to discover generalizable abstractions},
  author={Stengel-Eskin, Elias and Prasad, Archiki and Bansal, Mohit},
  journal={arXiv preprint arXiv:2401.16467},
  year={2024}
}

@article{wang2023voyager,
  title={Voyager: An open-ended embodied agent with large language models},
  author={Wang, Guanzhi and Xie, Yuqi and Jiang, Yunfan and Mandlekar, Ajay and Xiao, Chaowei and Zhu, Yuke and Fan, Linxi and Anandkumar, Anima},
  journal={arXiv preprint arXiv:2305.16291},
  year={2023}
}

@article{brockman2016openai,
  title={Openai gym},
  author={Brockman, Greg and Cheung, Vicki and Pettersson, Ludwig and Schneider, Jonas and Schulman, John and Tang, Jie and Zaremba, Wojciech},
  journal={arXiv preprint arXiv:1606.01540},
  year={2016}
}

@article{schulman2017ppo,
  title={Proximal policy optimization algorithms},
  author={Schulman, John and Wolski, Filip and Dhariwal, Prafulla and Radford, Alec and Klimov, Oleg},
  journal={arXiv preprint arXiv:1707.06347},
  year={2017}
}

@article{mnih2013dqn,
  title={Playing atari with deep reinforcement learning},
  author={Mnih, Volodymyr and Kavukcuoglu, Koray and Silver, David and Graves, Alex and Antonoglou, Ioannis and Wierstra, Daan and Riedmiller, Martin},
  journal={arXiv preprint arXiv:1312.5602},
  year={2013}
}

@article{ye2023deepaco,
  title={DeepACO: Neural-enhanced ant systems for combinatorial optimization},
  author={Ye, Haoran and Wang, Jiarui and Cao, Zhiguang and Liang, Helan and Li, Yong},
  journal={Advances in neural information processing systems},
  volume={36},
  pages={43706--43728},
  year={2023}
}

@article{dorigo2006ant,
  title={Ant colony optimization},
  author={Dorigo, Marco and Birattari, Mauro and Stutzle, Thomas},
  journal={IEEE computational intelligence magazine},
  volume={1},
  number={4},
  pages={28--39},
  year={2006},
  publisher={IEEE}
}

\clearpage
\typeout{PACE-APPENDIX-START-PAGE=\thepage}

\newcommand{\appplaceholder}[1]{%
  \begin{center}\fbox{\begin{minipage}[c][1.05in][c]{0.88\linewidth}\centering
  \small\textit{#1}\end{minipage}}\end{center}}

\section{Detailed Task Description}
\label{app:tasks}

This section defines the four tasks before specifying their evaluation settings. The control tasks require a heuristic policy, whereas the two routing tasks expose different algorithm-design interfaces for the same underlying optimization problem. The descriptions state the object being designed and the information available to it.

\subsection{Racing Car}

\begin{figure}[h!]
    \centering
    \includegraphics[width=\linewidth]{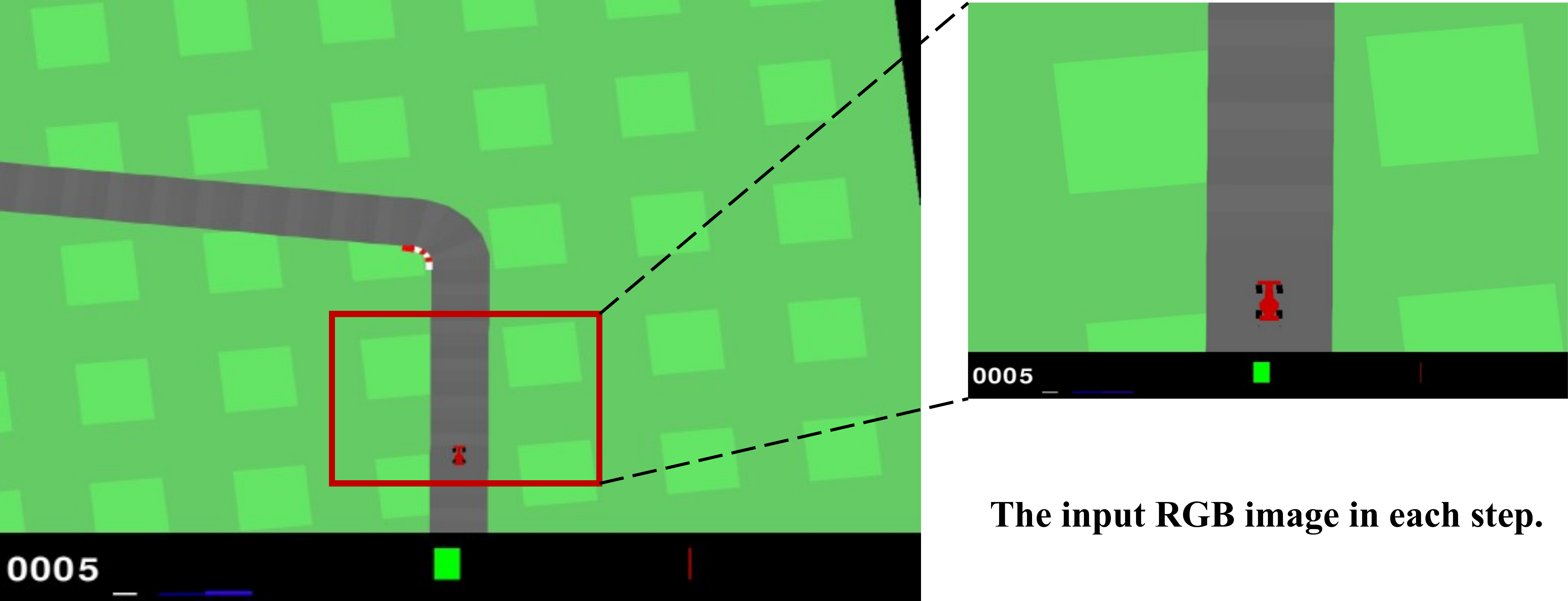}
    \caption{The visualization of task Racing Car.}
    \label{fig:racing_pic}
\end{figure}

Racing Car uses the continuous-action Gymnasium Racing Car environment \cite{brockman2016openai}, as shown in \figurename \ref{fig:racing_pic}. The target function receives an RGB observation, the current scalar speed, the previous action, and the previous observation. It returns a three-dimensional action consisting of steering, gas, and brake. The policy therefore combines visual perception with short-term action history. The objective is to control the car so that it follows the generated track and maximizes track coverage.

The task exposes several naturally separable policy components: extracting track direction, estimating a target speed, damping steering, and resolving the conflict between turning and acceleration. These components are useful examples for the EAP representation because an individual controller may be discarded while one of these local logic remains useful.

\subsection{Bipedal Walker}

\begin{figure}[h!]
    \centering
    \includegraphics[width=\linewidth]{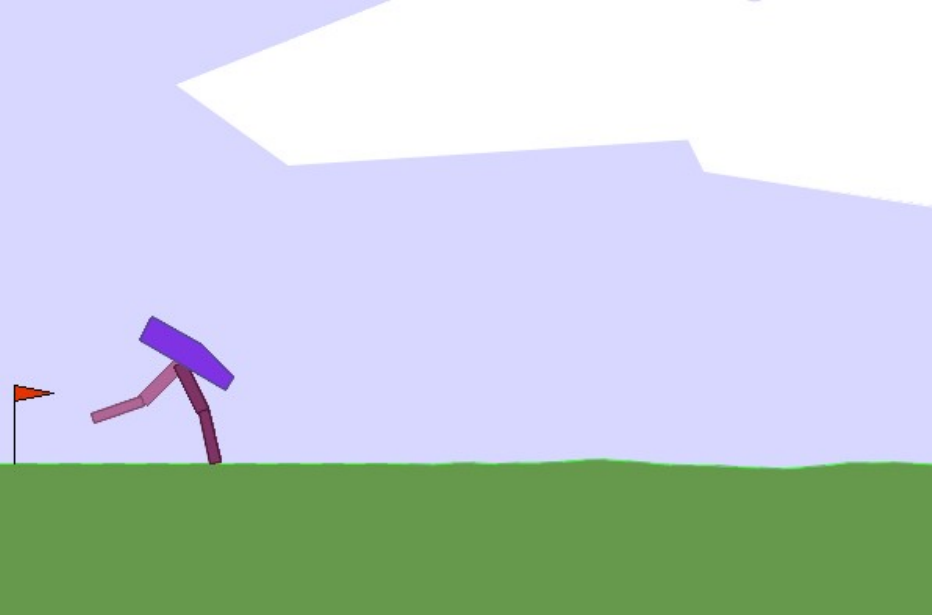}
    \caption{The visualization of task Bipedal Walker.}
    \label{fig:walker_pic}
\end{figure}

Bipedal Walker uses the continuous BipedalWalker-v3 environment \cite{brockman2016openai}, as shown in \figurename \ref{fig:walker_pic}. The target function receives a 24-dimensional observation, the previous four-dimensional action, and the previous observation. It returns four motor torques for the left hip, left knee, right hip, and right knee. The action is clipped to the valid range by the evaluator, and the objective is episodic return.

The observation contains hull orientation and angular velocity, horizontal and vertical velocity, the angles and angular velocities of both hip and knee joints, two foot-contact indicators, and ten lidar readings. The task is more strongly coupled than Racing Car because a local signal may help only when it is coordinated with support, swing, balance, and torque decisions elsewhere in the policy.

\subsection{TSP-Construct}

TSP-Construct evolves a function that receives the current node, the destination node, the set of unvisited nodes, and the distance matrix, and returns the next node. A complete tour is constructed by repeatedly calling this function until all nodes are visited and then returning to the start.

For both variants, let the complete graph be $G=(V,E)$ with $V=\{1,\ldots,n\}$ and distance matrix $D=(d_{ij})$. A tour is a permutation $\pi$ of $V$, and its cost is
\begin{equation}
L(\pi;D)=\sum_{t=1}^{n-1}d_{\pi_t,\pi_{t+1}}+d_{\pi_n,\pi_1}.
\end{equation}
The objective is to minimize $L$.

\subsection{TSP-ACO}

TSP-ACO evolves a function that maps a pairwise distance matrix to a finite heuristic matrix. The matrix is then consumed by a fixed ant-colony solver. The designed function is not the complete solver; it supplies the heuristic information used during tour construction.

TSP-ACO places the search boundary at the heuristic-matrix interface, whereas TSP-Construct places it at the local next-node decision. Thus, the two tasks share the same formal problem but evaluate different levels of algorithm design.

\section{Details of Evaluations \& Experiments}
\label{app:protocol-details}

This section specifies the datasets, evaluation procedures, and baseline configurations used in the experiments. All methods use the same evaluator and fixed instances for a given task, as shown in Table \ref{tab:app-unified-settings}. Test instances are evaluated only after the search selects the best-performing algorithm from the training performance.

\begin{table*}[t]
\centering
\small
\caption{Unified settings of evaluation algorithms, data generation, and experimential settings for control and routing tasks. Testing scores are computed using the best program selected by the training evaluator.}
\label{tab:app-unified-settings}
\begin{tabularx}{\textwidth}{p{0.11\textwidth} X p{0.22\textwidth} p{0.18\textwidth} X}
\toprule
\textbf{Task} & \textbf{Designed Algorithm} & \textbf{Environment / Data Setup} & \textbf{Training Protocol} & \textbf{Testing Protocol} \\
\midrule
\multicolumn{5}{l}{\textit{Continuous Control Tasks}} \\
\cmidrule{1-5}
Racing Car & Continuous policy from image and state history & Gym environment; fixed environment seeds & 4 fixed seeds; mean track coverage & 10 fixed seeds; training-best policy \\
\addlinespace[3pt]
Bipedal Walker & Continuous four-torque policy from state history & Gym environment; fixed environment seeds & 5 fixed seeds; mean episodic return & 10 fixed seeds; training-best policy \\
\midrule
\multicolumn{5}{l}{\textit{Routing Tasks}} \\
\cmidrule{1-5}
TSP-Construct & Next-node selection function & Uniform $[0,1]^2$, NumPy seed 2024 & 16 instances at $n=50$ & 64 fixed instances for each $n\in\{50,200,500,1000\}$ \\
\addlinespace[3pt]
TSP-ACO & Distance-to-heuristic matrix transformation & Uniform $[0,1]^2$, NumPy seed 1234  & 5 instances at $n=50$ & 64 fixed instances for each $n\in\{50,200,500,1000\}$ \\
\bottomrule
\end{tabularx}
\end{table*}

\subsection{LLM-based AAD Baselines}
\label{app:baseline-settings}

To ensure a fair comparison, all baseline methods, including EoH \cite{liu2024eoh}, ReEvo \cite{ye2024reevo}, MCTS-AHD \cite{zheng2025mctsahd}, and HsEvo \cite{dat2025hsevo}, retain their original operators while sharing the identical task evaluator and training instances as PACE. All hyperparameter configurations for these baselines follow the default settings reported in their original papers. The search budget for all algorithms is strictly normalized to $1,000$ complete-program evaluations. The detailed hyperparameter configurations for each baseline are summarized in Table~\ref{tab:baseline_settings}.

\begin{table}[htbp]
\centering
\small
\caption{Hyperparameter configurations for baseline algorithms.}
\label{tab:baseline_settings}
\begin{tabular}{lp{0.75\linewidth}}
\toprule
\textbf{Method} & \textbf{Hyperparameter Configuration} \\
\midrule
EoH & Population size = 20, samplers = 4, evaluators = 4 \\
ReEvo & Population size = 20, samplers = 4, evaluators = 4 \\
MCTS-AHD & Candidate pool size = 20, samplers = 4, evaluators = 4, parent selection = 2, $\alpha = 0.5$, $\lambda_0 = 0.1$ \\
HsEvo & Population size = 10 (initial = 30), samplers = 8, evaluators = 8, mutation rate = 0.5, HMS = 5, HMCR = 0.7, PAR = 0.5, bandwidth = 0.2, harmony iterations = 5 \\
\bottomrule
\end{tabular}
\end{table}

\subsection{MLES}


To adapt the multimodal baseline MLES for our benchmark, we utilize its proposed $e1$, $e2$, $m1_M$, and $m2_M$ operators with a population size of 16, two parents, eight samplers, and eight evaluators. We employ GPT-4o-mini at temperature 1.0 and constrain the search budget to 1,000 model calls, adjusting from the 2,000 calls in the original paper. For both control tasks, MLES receives standard rendered behavioral evidence under a cold-start condition.

\begin{figure}[h!]
    \centering
    \includegraphics[width=\linewidth]{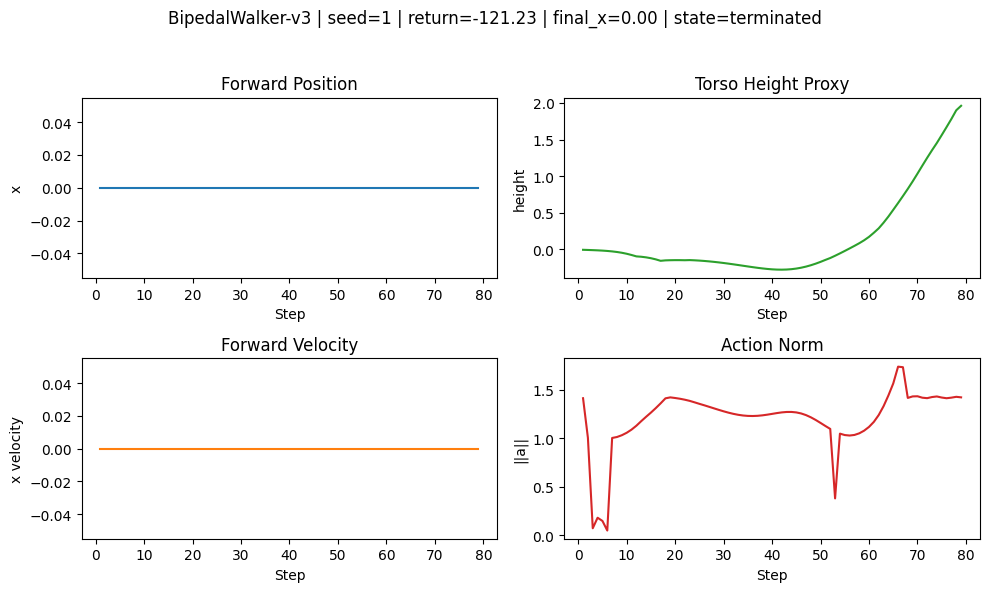}
    \caption{The visualization of task Bipedal Walker.}
    \label{fig:mles_walker}
\end{figure}

For Bipedal Walker, which relies on a 24-dimensional continuous state vector rather than native image inputs, standard environment frames fail to reflect the underlying state dynamics. To ensure a fair comparison under MLES's multimodal requirement, we supply its interface with a trajectory visualization that formats vector dynamics into an image representation. Specifically, the evaluator extracts the least successful reference episode and plots its trajectory into a 4-panel figure, accompanied by twelve sampled numerical records. As shown in Figure~\ref{fig:mles_walker}, the top banner reports episode-level metadata (total return, final displacement, and termination status), while the four subplots track forward position, torso height proxy, forward velocity, and action norm ($\|a\|_2$). This setup provides MLES with necessary behavioral evidence to inspect failure modes such as balance loss and forward stagnation without altering its multimodal framework.

\subsection{PACE}

Table~\ref{tab:app-pace-settings} details the experimental setup for the PACE framework. Following \cite{liu2024eoh}, the population size is set to $20$. In each generation, at most $1$ EAP proposal is generated. To control search complexity, a unified parameter $k=3$ bounds both the maximum number of EAPs per algorithm and the maximum structural generation attempts.

\begin{table}[t]
\centering
\caption{Experimental settings of PACE.}
\begin{tabular}{p{0.54\columnwidth}p{0.40\columnwidth}}
\toprule
PACE setting & Value \\
\midrule
Population size & 20 \\
Maximum EAPs per algorithm, $k$ & 3 \\
EAP Generation / Extraction proposals & At most 1 per generation \\
Structural generation attempts, $k$ & 3 \\
\bottomrule
\end{tabular}
\label{tab:app-pace-settings}
\end{table}

\begin{table*}[h!]
\centering
\small
\caption{Configurations and execution status for non-AAD classical and learned baselines.}
\label{tab:app-non-aad}
\begin{tabular}{p{0.20\linewidth} p{0.14\linewidth} p{0.45\linewidth} p{0.13\linewidth}}
\toprule
\textbf{Baseline} & \textbf{Task} & \textbf{Hyperparameter Configuration} & \textbf{Status} \\
\midrule
PPO (Racing Car) & Racing Car & Conv policy, LR = $2.5\times10^{-4}$, rollout = 512, batch = 512, 3 epochs, $\gamma = 0.99$, $\lambda = 0.95$, 4,000 resets (4 seeds) & Formal runs \\
\addlinespace[3pt]
PPO (Bipedal Walker) & Bipedal Walker & MLP policy, LR = $3\times10^{-4}$, rollout = 2048, batch = 256, 10 epochs, $\gamma = 0.99$, $\lambda = 0.95$, 4,000 resets (5 seeds) & Formal runs \\
\addlinespace[3pt]
Greedy Construct & TSP-Construct & Start node 0, nearest unvisited node selection & 64-instance test \\
\addlinespace[3pt]
ACO / DeepACO & TSP-ACO & 30 ants, 100 iterations, official pretrained checkpoints & 64-instance test \\
\bottomrule
\end{tabular}
\end{table*}

\subsection{Learned and Classical Baselines}

To evaluate classical and learned non-AAD baselines, we include Proximal Policy Optimization (PPO) \cite{schulman2017ppo} for continuous control tasks, alongside Greedy Construction and Ant Colony Optimization (ACO) \cite{dorigo2006ant, ye2023deepaco} variants for routing problems. All learned baselines are evaluated under standardized execution protocols.

For continuous control, PPO is trained with a total budget of 4,000 completed environment resets, allocated as 1,000 resets per seed across four seeds. Although Racing Car permits discrete action adaptation, preliminary DQN \cite{mnih2013dqn} runs served solely as internal sanity checks and are excluded from formal evaluation. Meanwhile, Bipedal Walker operates on a four-dimensional continuous action space, precluding direct comparison with standard DQN algorithms.

For TSP-Construct, Greedy Construction deterministically selects the nearest unvisited node starting from node zero. For TSP-ACO, evaluations of learning-based routing baselines such as DeepACO strictly adhere to official pretrained checkpoints and inference protocols, thereby avoiding performance degradation caused by uncalibrated models.

The detailed hyperparameter configurations and appendix completion status for all non-AAD baselines are summarized in Table~\ref{tab:app-non-aad}.

\begin{table*}[h!]
\centering
\scriptsize
\caption{Representative best-program descriptions. Names are shown to make the EAP transfer traceable; they are not additional task-specific primitives.}
\begin{tabular}{p{0.14\linewidth}p{0.32\linewidth}p{0.42\linewidth}}
\toprule
Task & EAP calls & Algorithm description \\
\midrule
Racing Car & \url{compute_adjusted_action}; \url{compute_steering_and_throttle}; \url{compute_throttle_and_brake} & Estimate track direction and speed, damp steering changes, then resolve throttle and braking from current and previous observations. \\
Bipedal Walker & \url{calculate_dynamic_torques}; \url{compute_torque}; \url{extract_velocity_features} & Adapt motor torques from velocity features and current state while retaining a recovery rule for unstable body motion. \\
TSP-ACO & \url{compute_attraction_score_matrix}; \url{compute_complex_attraction_matrix}; \url{compute_quadratic_attraction_matrix} & Combine distance-sensitive, nonlinear, and quadratic edge preferences before ACO execution. \\
TSP-Construct & \url{compute_average_neighbor_distance}; \url{compute_node_diversity}; \url{compute_priority_score} & Score unvisited nodes using proximity, neighborhood diversity, and cumulative distance penalties. \\
\bottomrule
\end{tabular}
\label{tab:app-best-programs}
\end{table*}


\subsection{Token Usage}
\label{app:run-results}

Table~\ref{tab:app-token-usage} reports total API token consumption for the exact three runs underlying the results in the main paper. Total usage includes prompt and completion tokens. For PACE, this includes calls for both complete programs and EAP discovery. For the baselines, it includes candidate generation and method-specific auxiliary calls. MLES is evaluated only on the two control tasks. PPO, Greedy Construction, ACO, and DeepACO do not call an LLM and are therefore omitted from the table.

\begin{table*}[h]
\centering
\caption{LLM token consumption for the search runs reported in the main paper. Values are mean $\pm$ standard deviation over three independent runs.}
\label{tab:app-token-usage}
\begin{tabular}{@{} l cccc @{}}
\toprule
& \multicolumn{4}{c}{Total tokens ($\times 10^6$), GPT-4o-mini} \\
\cmidrule(l){2-5}
Method & Racing Car & Bipedal Walker & TSP-ACO & TSP-Construct \\
\midrule
PACE        & $\phantom{1}4.88 \pm 0.81$ & $\phantom{1}3.66 \pm 0.71$ & $1.85 \pm 0.33$ & $1.72 \pm 0.32$ \\
EoH         & $\phantom{1}2.58 \pm 0.15$ & $\phantom{1}1.73 \pm 0.01$ & $0.85 \pm 0.06$ & $0.84 \pm 0.04$ \\
ReEvo       & $\phantom{1}4.83 \pm 0.15$ & $\phantom{1}3.68 \pm 0.28$ & $2.25 \pm 0.04$ & $1.76 \pm 0.01$ \\
MCTS-AHD    & $\phantom{1}4.94 \pm 0.44$ & $\phantom{1}3.88 \pm 0.42$ & $1.99 \pm 0.17$ & $1.97 \pm 0.12$ \\
HsEvo       & $\phantom{1}2.72 \pm 0.29$ & $\phantom{1}2.04 \pm 0.26$ & $1.40 \pm 0.15$ & $1.39 \pm 0.09$ \\
MLES (cold) & $11.22 \pm 2.91$           & $14.69 \pm 0.40$           & --              & --              \\
\bottomrule
\end{tabular}
\end{table*}

\section{Detailed Methodology}
\label{app:methodology}

This section gives the prompts, representative programs, and complete PACE procedure. The prompts define each operation, while the parsed call set is used to verify its structural contract before evaluation.

\subsection{Prompts}
\label{app:prompts}

This section reports the stable instruction layer of each prompt. At run time,
PACE fills the shaded input fields with the task description, target template,
parent programs, and EAP implementations. Operator prompts are shown in blue,
whereas EAP-generation prompts are shown in green. This separation mirrors
their roles in the search: the former produce complete algorithms and the
latter produce reusable functions.

\definecolor{pacePromptBlue}{HTML}{245B8F}
\definecolor{pacePromptBlueLight}{HTML}{F2F6FA}
\definecolor{pacePromptGreen}{HTML}{2F6B59}
\definecolor{pacePromptGreenLight}{HTML}{F1F7F4}
\definecolor{pacePromptInput}{HTML}{E8EDF2}
\definecolor{paceCodeBlue}{HTML}{1F4E79}
\definecolor{paceCodeGreen}{HTML}{2E6B4F}
\definecolor{paceCodeRed}{HTML}{9A3B3B}
\definecolor{paceCodeGray}{HTML}{6B7280}
\definecolor{paceCodeBack}{HTML}{F7F8FA}

\newtcolorbox{paceoperatorprompt}[1]{
  enhanced, breakable, sharp corners, boxrule=0.75pt,
  colframe=pacePromptBlue, colback=pacePromptBlueLight,
  colbacktitle=pacePromptBlue, coltitle=white,
  fonttitle=\sffamily\bfseries, title={#1},
  left=7pt, right=7pt, top=6pt, bottom=6pt,
  before skip=7pt, after skip=9pt,
  borderline west={2.2pt}{0pt}{pacePromptBlue}
}
\newtcolorbox{paceeapprompt}[1]{
  enhanced, breakable, sharp corners, boxrule=0.75pt,
  colframe=pacePromptGreen, colback=pacePromptGreenLight,
  colbacktitle=pacePromptGreen, coltitle=white,
  fonttitle=\sffamily\bfseries, title={#1},
  left=7pt, right=7pt, top=6pt, bottom=6pt,
  before skip=7pt, after skip=9pt,
  borderline west={2.2pt}{0pt}{pacePromptGreen}
}
\newcommand{\promptfield}[1]{%
  \par\smallskip\noindent
  \textcolor{paceCodeGray}{\sffamily\bfseries #1}%
  \hspace{0.5em}\ignorespaces 
}

\newcommand{\promptinput}[1]{%
  \begingroup
  \setlength{\fboxsep}{2pt}
  \colorbox{pacePromptInput}{\small\ttfamily #1}
  \endgroup
}

\renewcommand{\promptfield}[1]{%
  \par\smallskip\noindent\raggedright
  \textcolor{paceCodeGray}{\sffamily\bfseries #1}\hspace{0.5em}\ignorespaces
}

\renewcommand{\promptinput}[1]{%
  \begingroup
  \setlength{\fboxsep}{1.5pt}
  \colorbox{pacePromptInput}{\small\ttfamily #1}%
  \endgroup
}

\tcbset{every box/.append style={halign=left}}


\lstdefinestyle{pacepython}{
  language=Python,
  basicstyle=\scriptsize\ttfamily,
  keywordstyle=\color{paceCodeBlue}\bfseries,
  commentstyle=\color{paceCodeGray}\itshape,
  stringstyle=\color{paceCodeRed},
  identifierstyle=\color{black},
  numbers=left,
  numberstyle=\tiny\color{paceCodeGray},
  numbersep=8pt,
  stepnumber=1,
  backgroundcolor=\color{paceCodeBack},
  frame=single,
  rulecolor=\color{pacePromptBlue!65},
  framerule=0.6pt,
  framesep=6pt,
  xleftmargin=2.2em,
  framexleftmargin=1.7em,
  breaklines=true,
  breakatwhitespace=false,
  showstringspaces=false,
  keepspaces=true,
  columns=fullflexible,
  tabsize=4,
  captionpos=b,
  aboveskip=8pt,
  belowskip=10pt
}

\begin{paceoperatorprompt}{I1: Complete Algorithm Initialization}
\small
\promptfield{Inputs} \promptinput{TASK DESCRIPTION}, \promptinput{FUNCTION TEMPLATE}
\promptfield{Instruction} First describe the new algorithm and its main steps in one sentence inside braces. Then implement the target function without changing its signature. Follow the declared input and return contracts.
\promptfield{Output contract} Return the description followed by one complete target-function implementation. Do not provide additional explanation.
\end{paceoperatorprompt}

\begin{paceoperatorprompt}{P1: EAP Insertion}
\small
\promptfield{Inputs} \promptinput{TASK DESCRIPTION}, \promptinput{FUNCTION TEMPLATE}, \promptinput{PARENT}, \promptinput{PROTECTED EAPs}, \promptinput{FOCUS EAP}
\promptfield{Instruction} Insert the focus EAP into a complete offspring while preserving the mechanisms of the parent.
\promptfield{Structural contract} The offspring must call \texttt{FOCUS} and every protected EAP called by the parent. It must not merely wrap or rename the parent.
\promptfield{Output contract} Return a one-sentence description in braces followed by the complete target function. Do not output EAP definitions.
\end{paceoperatorprompt}

\begin{paceoperatorprompt}{P2: EAP Replacement}
\small
\promptfield{Inputs} \promptinput{TASK DESCRIPTION}, \promptinput{FUNCTION TEMPLATE}, \promptinput{PARENT}, \promptinput{PROTECTED EAPs}, \promptinput{TARGET EAP}, \promptinput{REPLACEMENT EAP}
\promptfield{Instruction} Replace the target EAP with the replacement EAP and adapt the complete algorithm so that the replacement is used meaningfully.
\promptfield{Structural contract} The offspring must call \texttt{REPLACEMENT}, must not call \texttt{TARGET}, and must preserve every other protected EAP.
\promptfield{Output contract} Return the complete target function without EAP definitions.
\end{paceoperatorprompt}

\begin{paceoperatorprompt}{P3: EAP-Preserving Refinement}
\small
\promptfield{Inputs} \promptinput{TASK DESCRIPTION}, \promptinput{FUNCTION TEMPLATE}, \promptinput{PARENT}, \promptinput{PARENT EAP IMPLEMENTATIONS}
\promptfield{Instruction} Refine how the complete algorithm invokes, scales, combines, or conditions its existing EAPs. Thresholds and fallback behavior may also be changed.
\promptfield{Structural contract} The offspring must call exactly \promptinput{PARENT EAP NAMES}; no EAP may be added or removed.
\promptfield{Output contract} Return the refined complete target function without EAP definitions.
\end{paceoperatorprompt}

\begin{paceoperatorprompt}{P4: EAP-Aware Crossover}
\small
\promptfield{Inputs} \promptinput{TASK DESCRIPTION}, \promptinput{FUNCTION TEMPLATE}, \promptinput{PARENT 1}, \promptinput{PARENT 2}, \promptinput{EAP IMPLEMENTATIONS}
\promptfield{Instruction} Create one complete offspring that combines meaningful mechanisms from both parents.
\promptfield{Structural contract} The offspring must call every designated EAP, including at least one inherited from each parent, and may call at most $k$ EAPs in total.
\promptfield{Output contract} Return one description and the complete target function without EAP definitions.
\end{paceoperatorprompt}

\begin{paceeapprompt}{E1: Initialing EAP}
\small
\promptfield{Inputs} \promptinput{TASK DESCRIPTION}, \promptinput{FUNCTION TEMPLATE}
\promptfield{Instruction} Design reusable local evaluators, feature extractors, or compact mid-level modules that may support complete algorithms for the task.
\promptfield{Function contract} Each function must be narrower than the target function, self-contained, and restricted to inputs in the target signature. It must neither call another EAP nor reimplement the target function.
\promptfield{Output contract} Return Python function definitions only.
\end{paceeapprompt}

\begin{paceeapprompt}{E2: EAP Extraction}
\small
\promptfield{Inputs} \promptinput{TASK DESCRIPTION}, \promptinput{History Best Algorithm}
\promptfield{Instruction} Identify one complex internal computation that contributes to the candidate and refactor it into a reusable function.
\promptfield{Function contract} The function must be standalone, smaller than the complete algorithm, restricted to target-function inputs, and unable to call another EAP.
\promptfield{Output contract} Return exactly one function definition and no explanation.
\end{paceeapprompt}

\begin{paceeapprompt}{E3: EAP Generation}
\small
\promptfield{Inputs} \promptinput{TASK}, \promptinput{TARGET SIGNATURE}, \promptinput{TOP-$k$ EAP SUMMARIES}
\promptfield{Instruction} Generate one self-contained function whose capability differs from the current EAPs.
\promptfield{Novelty contract} Do not produce a superficial variant through symbol renaming, constant or exponent changes, or an equivalent transformation. Use only target-function inputs.
\promptfield{Output contract} Return exactly one function definition and no explanation.
\end{paceeapprompt}

\subsection{Representative best programs}

The following compact descriptions identify the best PACE programs used for the four task examples, summary ara shown in Table \ref{tab:app-best-programs}. They describe the algorithmic role of the EAP calls. The complete source and exact prompt are retained in the corresponding run directory.

Listings~\ref{lst:app-best-Racing Car}--\ref{lst:app-best-tspconstruct}
contain the exact complete program and EAP implementations used by the four
representative solutions. The algorithm description, training score, and
search evaluation are included in the header of each listing. Keeping the EAP
implementations beside the complete function makes every external call in the
selected program explicit.

\onecolumn

\begin{lstlisting}[
  style=pacepython,
  morekeywords={[2]{compute_adjusted_action,compute_steering_and_throttle,compute_throttle_and_brake}},
  keywordstyle={[2]\color{paceCodeGreen}\bfseries},
  morekeywords={[3]{choose_action}},
  keywordstyle={[3]\color{paceCodeBlue}\bfseries},
  caption={Selected PACE program for Racing Car. Green identifiers denote EAPs and blue denotes the complete target function.},
  label={lst:app-best-Racing Car}
]
# Algorithm description: Estimate track direction and speed, damp steering changes, then coordinate steering, throttle, and braking from current and previous observations.

# Training score: 100.000000000

# Search evaluation: 176

# Executable Algorithmic Primitives

def compute_adjusted_action(car_speed: float, action: np.ndarray, pre_action: np.ndarray) -> np.ndarray:
    """Adjust the action array based on speed and previous actions; controls gas and brake behavior."""
    # Release brake if the previous action was braking
    if pre_action[2] > 0:  # Previously braking
        action[2] = 0  # Release brake if it was previously pressing

    # Reduce gas if turning sharply and speed is high to maintain control
    if abs(action[0]) > 0.5 and car_speed > 20:  # If turning sharply and speed is high
        action[1] = min(action[1], 0.5)  # Limit gas to enhance control

    return action

def compute_steering_and_throttle(observation: np.ndarray, car_speed: float, pre_action: np.ndarray) -> np.ndarray:
    """Compute steering and throttle based on the current observation and previous action; returns an action array."""
    MAX_STEERING = 1.0
    MAX_GAS = 0.6
    MIN_GAS = 0.1  # never zero to avoid total stop
    SPEED_LIMIT_SHARP_TURN = 15.0  # speed threshold for reducing gas in sharp turns
    SHARP_TURN_ANGLE = 0.5  # threshold in steering value to consider turn sharp

    H, W, _ = observation.shape

    # Step 1: Identify pixels of car, track, off-track grass, and curbs by color thresholds
    R = observation[:, :, 0]
    G = observation[:, :, 1]
    B = observation[:, :, 2]

    # Mask for car, track, grass, and curbs
    car_mask = (R > 190) & (R < 215) & (G < 20) & (B < 20)
    track_mask = (np.abs(R - 102) < 20) & (np.abs(G - 102) < 20) & (np.abs(B - 102) < 20)
    grass_mask = (np.abs(R - 102) < 40) & (np.abs(G - 204) < 50) & (np.abs(B - 102) < 40)
    curb_mask = ((R > 240) & (G < 20) & (B < 20)) | ((R > 240) & (G > 240) & (B > 240))

    # Step 2: Locate car centroid    
    car_positions = np.argwhere(car_mask)
    if car_positions.shape[0] == 0:
        # Fallback if car not found: 
        return np.array([0.0, max(MIN_GAS, pre_action[1] * 0.8), 0.0])

    car_y, car_x = car_positions.mean(axis=0)

    # Step 3: Extract the track region ahead of the car to estimate curvature and direction
    look_ahead_height = 20
    look_ahead_width = 40
    patch_top = int(max(car_y - look_ahead_height - 5, 0))
    patch_bottom = int(max(car_y - 5, 0))
    patch_left = int(max(car_x - look_ahead_width // 2, 0))
    patch_right = int(min(car_x + look_ahead_width // 2, W-1))

    track_patch = track_mask[patch_top:patch_bottom, patch_left:patch_right]

    if track_patch.size == 0 or np.sum(track_patch) < 10:
        return np.array([pre_action[0] * 0.9, max(MIN_GAS, pre_action[1] * 0.7), 0.0])

    cols = np.arange(track_patch.shape[1])
    col_weights = track_patch.sum(axis=0)
    if col_weights.sum() == 0:
        return np.array([0.0, MIN_GAS, 0.3])

    center_x_in_patch = np.sum(cols * col_weights) / col_weights.sum()
    offset = (center_x_in_patch - track_patch.shape[1] / 2) / (track_patch.shape[1] / 2)  

    steer = np.clip(offset, -MAX_STEERING, MAX_STEERING)

    # Step 6: Adjust gas based on turn and speed
    sharp_turn = (abs(steer) > SHARP_TURN_ANGLE) or (np.sum(curb_mask[patch_top:patch_bottom, patch_left:patch_right]) / track_patch.size > 0.15)
    gas, brake = 0.0, 0.0

    if sharp_turn:
        if car_speed > SPEED_LIMIT_SHARP_TURN:
            gas = MIN_GAS
            brake = min(0.6, 0.5 + 0.5 * (car_speed - SPEED_LIMIT_SHARP_TURN)/20)
        else:
            gas = MAX_GAS * 0.5
    else:
        gas = MAX_GAS if car_speed < 30.0 else max(MIN_GAS, MAX_GAS * (1 - (car_speed - 30)/40))

    return np.array([steer, gas, brake])

def compute_throttle_and_brake(car_speed: float, action: np.ndarray, pre_action: np.ndarray) -> np.ndarray:
    """Computes adjusted throttle and brake values based on car speed and previous actions; returns an array of [gas, brake]."""
    MAX_GAS = 0.6
    MIN_GAS = 0.1  # never zero to avoid total stop

    # Adjusting throttle to ensure the car does not come to a full stop
    if action[1] < MIN_GAS:
        action[1] = MIN_GAS

    # Avoiding excessive braking if previous action was also braking
    if pre_action[2] > 0.5:
        action[2] = min(0.5, pre_action[2] * 0.5)

    # Maintaining throttle if speed is excessively low
    if car_speed < 7.0:  # Condition updated for better responsiveness
        action[1] = max(action[1], 0.4)

    return action

# Complete algorithm

import numpy as np
import cv2
def choose_action(observation, car_speed, pre_action, pre_observation):
    """
    Determine the next action for the Car Racing agent.
    This function takes into account the current state (observation and speed), the previous action, and the previous observation.

    Notes:
    - The car in this environment is a powerful rear-wheel-drive vehicle. Avoid accelerating while turning sharply,
      as this can easily lead to loss of control.
    - Occasionally, track segments (e.g., after a U-turn) may appear in the observation but are not part of the immediate drivable path. These should be distinguished to avoid premature or incorrect decisions.
    - Avoid coming to a complete stop, as this may prevent the car from finishing the race.

    Args:
        observation (np.ndarray): The current state observed by the agent, represented as a 96x96 RGB image
                                   of the car and race track from a top-down view (shape: (96, 96, 3)).

        car_speed (float): The current speed of the car.

        pre_action (np.ndarray): The action taken by the agent in the previous step, represented as a
                                  3-element array.

        pre_observation (np.ndarray): The observation received when the previous action was taken. It has the same shape and format as `observation` (i.e., a 96x96 RGB image).
        
    Returns:
        np.ndarray: The action selected by the agent for the next step, represented as an array of shape (3,) where:
                    - Index 0: Steering, where -1 is full left, +1 is full right (range: [-1, 1]).
                    - Index 1: Gas, (range: [0, 1]).
                    - Index 2: Braking, (range: [0, 1]).
    """
    
    action = compute_steering_and_throttle(observation, car_speed, pre_action)
    action = compute_adjusted_action(car_speed, action, pre_action)
    action = compute_throttle_and_brake(car_speed, action, pre_action)

    return action
\end{lstlisting}

\begin{lstlisting}[
  style=pacepython,
  morekeywords={[2]{calculate_dynamic_torques,compute_torque,extract_velocity_features}},
  keywordstyle={[2]\color{paceCodeGreen}\bfseries},
  morekeywords={[3]{choose_action}},
  keywordstyle={[3]\color{paceCodeBlue}\bfseries},
  caption={Selected PACE program for Bipedal Walker. Green identifiers denote EAPs and blue denotes the complete target function.},
  label={lst:app-best-bipedal}
]
# Algorithm description: Adapt four motor torques from velocity features and current state while retaining recovery behavior for unstable body motion.

# Training score: 143.110807357

# Search evaluation: 634

# Executable Algorithmic Primitives

def calculate_dynamic_torques(observation: np.ndarray, last_action: np.ndarray, prev_observation: np.ndarray) -> np.ndarray:
    """Computes dynamic motor torques for the Bipedal Walker based on observed states; returns four torques for [left hip, left knee, right hip, right knee] in [-1, 1]."""
    
    # Extract observation features for easier referencing
    hull_angle = observation[0]
    hull_velocity = observation[1:3]
    joints = observation[3:7]
    contacts = observation[7:11]
    lidar_data = observation[11:21]
    
    # Calculate balance and stability torques
    balance_torque = -np.clip(hull_angle * 0.5, -1, 1)  # Small corrective torque based on hull angle
    stability_adjustments = np.zeros(4)
    
    # Adjust for the joint states and their contacts
    for i in range(2):
        if contacts[i] > 0:  # If there's contact, provide more torque to stabilize
            stability_adjustments[i * 2] = np.clip(-joints[i * 2] * 0.5, -1, 1)  # Left hip/knee stabilization
            stability_adjustments[i * 2 + 1] = np.clip(-joints[i * 2 + 1] * 0.5, -1, 1)  # Right hip/knee stabilization
    
    # Forward motion control
    forward_torque = np.clip(hull_velocity[0] * 0.1, -1, 1)  # Simple forward speed control adjustment

    # Combine all torques into a final control signal
    torques = np.array([
        balance_torque + stability_adjustments[0] + forward_torque,
        stability_adjustments[1],
        balance_torque + stability_adjustments[2] + forward_torque,
        stability_adjustments[3]
    ])
    
    return np.clip(torques, -1, 1)  # Ensure torques are within the allowed range

def compute_torque(hull_angle: float, joint_angle: float, joint_velocity: float, last_action: float) -> float:
    """Calculates torque for a joint based on hull angle, joint angle, joint velocity, and last action."""
    torque = (
        -1.5 * hull_angle  # Influence from hull orientation
        - 0.6 * joint_angle  # Penalty for joint's angle deviation
        + (0.5 * last_action if np.abs(joint_angle) < 0.25 else 0.0)  # Prior action influence if angle is small
        + (0.15 * joint_velocity if joint_velocity < 0 else 0.0)  # Consider velocity if moving backward
    )
    return torque

def extract_velocity_features(observation: np.ndarray) -> np.ndarray:
    """Extract velocity-related features: horizontal and vertical velocities."""
    return observation[2:4]  # Returns a 2D array with horizontal and vertical velocities.

# Complete algorithm

import numpy as np
def choose_action(observation: np.ndarray, last_action: np.ndarray, prev_observation: np.ndarray) -> np.ndarray:
    """
    Select the four motor torques for BipedalWalker-v3.

    Args:
        observation: Current 24-dimensional state. Entries are ordered as:
            0 hull angle, 1 hull angular velocity, 2 horizontal velocity,
            3 vertical velocity, 4 left hip angle, 5 left hip angular velocity,
            6 left knee angle, 7 left knee angular velocity, 8 left foot contact,
            9 right hip angle, 10 right hip angular velocity, 11 right knee angle,
            12 right knee angular velocity, 13 right foot contact, and
            14:24 normalized lidar range readings in the environment's fixed ray order.
            Joint angles and velocities use the environment's normalized coordinates;
            contact entries are binary.
        last_action: The previous four torques in left-hip, left-knee, right-hip,
            right-knee order.
        prev_observation: The observation from the preceding environment step,
            with the same layout as observation.

    Returns:
        A finite array of four torques in [-1, 1], in left-hip, left-knee,
        right-hip, right-knee order.
    """
    # {The algorithm enhances stability and control by intelligently adapting motor torques 
    # based on observed states while effectively managing recovery strategies.}
    hull_angle = observation[0]

    # Joint angles and velocities extraction
    left_hip_angle, left_hip_velocity = observation[4], observation[5]
    left_knee_angle, left_knee_velocity = observation[6], observation[7]
    right_hip_angle, right_hip_velocity = observation[9], observation[10]
    right_knee_angle, right_knee_velocity = observation[11], observation[12]

    # Extracting velocity features
    horizontal_velocity, vertical_velocity = extract_velocity_features(observation)

    # Compute dynamic torques for balance and recovery adjustments
    dynamic_torques = calculate_dynamic_torques(observation, last_action, prev_observation)

    left_hip_torque = dynamic_torques[0] + compute_torque(hull_angle, left_hip_angle, left_hip_velocity, last_action[0])
    left_knee_torque = dynamic_torques[1] + (
        -0.7 * left_knee_angle
        + 0.4 * last_action[1] * (1 if left_knee_angle < 0.2 else 0)
        + 0.1 * left_knee_velocity * (1 if left_knee_velocity < 0 else 0)
    )
    right_hip_torque = dynamic_torques[2] + compute_torque(-hull_angle, right_hip_angle, right_hip_velocity, last_action[2])
    right_knee_torque = dynamic_torques[3] + (
        -0.7 * right_knee_angle
        + 0.4 * last_action[3] * (1 if right_knee_angle < 0.2 else 0)
        + 0.1 * right_knee_velocity * (1 if right_knee_velocity < 0 else 0)
    )

    torques = np.array([left_hip_torque, left_knee_torque, right_hip_torque, right_knee_torque])
    torques = np.clip(torques, -1, 1)

    return torques
\end{lstlisting}

\begin{lstlisting}[
  style=pacepython,
  morekeywords={[2]{compute_average_neighbor_distance,compute_node_diversity,compute_priority_score}},
  keywordstyle={[2]\color{paceCodeGreen}\bfseries},
  morekeywords={[3]{select_next_node}},
  keywordstyle={[3]\color{paceCodeBlue}\bfseries},
  caption={Selected PACE program for TSP-Construct. Green identifiers denote EAPs and blue denotes the complete target function.},
  label={lst:app-best-tspconstruct}
]
# Algorithm description: Score unvisited nodes by proximity, neighborhood diversity, and cumulative distance penalties.

# Training score: 6.103043479

# Search evaluation: 850

# Executable Algorithmic Primitives

def compute_average_neighbor_distance(current_node: int, unvisited_nodes: np.ndarray, distance_matrix: np.ndarray) -> float:
    """Computes the average distance from the current node to all unvisited nodes; helps assess overall route efficiency."""
    if len(unvisited_nodes) == 0:
        return float('inf')  # Return infinity if there are no unvisited nodes

    total_distance = 0
    for node in unvisited_nodes:
        total_distance += distance_matrix[current_node][node]
    
    average_distance = total_distance / len(unvisited_nodes)
    return average_distance

def compute_node_diversity(current_node: int, unvisited_nodes: np.ndarray, distance_matrix: np.ndarray) -> np.ndarray:
    """Calculates a diversity score for each unvisited node based on how far they are from each other, promoting exploration of less clustered nodes."""
    diversity_scores = np.zeros(len(unvisited_nodes))
    
    for i, node in enumerate(unvisited_nodes):
        distances_to_others = distance_matrix[node][unvisited_nodes]
        diversity_scores[i] = np.mean(distances_to_others)

    return diversity_scores

def compute_priority_score(current_node: int, node: int, destination_node: int, cumulative_distance: float, distance_matrix: np.ndarray) -> float:
    """Computes a priority score for a node based on its distance from the current node and its distance to the destination."""
    distance_to_destination = distance_matrix[node][destination_node]
    distance_from_current = distance_matrix[current_node][node]
    
    # Enhanced heuristic with a more adaptive penalty based on the ratio of distances
    heuristic_penalty = (np.log1p(distance_to_destination) / (1 + np.log1p(distance_to_destination))) if distance_to_destination > 0 else 0
    
    # Updated priority score calculations with more emphasis on proximity to the current node
    priority_score = (1.5 / (distance_from_current + 1)) + heuristic_penalty - (0.3 * cumulative_distance)
    
    return priority_score

# Complete algorithm

import numpy as np
def select_next_node(current_node: int, destination_node: int, unvisited_nodes: np.ndarray, distance_matrix: np.ndarray) -> int:
    """
    {This algorithm emphasizes a synergy between proximity, diversity, and cumulative distance penalties to optimize route selection.}
    """
    cumulative_distance_from_start = 0  # to be updated in the overall algorithm
    best_node = None
    best_priority_score = -np.inf

    average_neighbor_distance = compute_average_neighbor_distance(current_node, unvisited_nodes, distance_matrix)
    diversity_scores = compute_node_diversity(current_node, unvisited_nodes, distance_matrix)

    for index, node in enumerate(unvisited_nodes):
        distance_from_current = distance_matrix[current_node][node]
        priority_score = compute_priority_score(current_node, node, destination_node, cumulative_distance_from_start, distance_matrix)

        exploration_factor = (diversity_scores[index] ** 1.5) / (1 + np.sqrt(distance_from_current))  # Encourage exploration of diverse nodes with squared scaling
        proximity_adjustment = (distance_from_current - average_neighbor_distance) * 0.05  # Reduced balancing impact for more leniency
        
        modified_score = priority_score + exploration_factor - proximity_adjustment

        if modified_score > best_priority_score:
            best_priority_score = modified_score
            best_node = node

    return best_node if best_node is not None else unvisited_nodes[0]
\end{lstlisting}

\begin{lstlisting}[
  style=pacepython,
  morekeywords={[2]{compute_attraction_score_matrix,compute_complex_attraction_matrix,compute_quadratic_attraction_matrix}},
  keywordstyle={[2]\color{paceCodeGreen}\bfseries},
  morekeywords={[3]{heuristics}},
  keywordstyle={[3]\color{paceCodeBlue}\bfseries},
  caption={Selected PACE program for TSP-ACO. Green identifiers denote EAPs and blue denotes the complete target function.},
  label={lst:app-best-tspaco}
]
# Algorithm description: Combine distance-sensitive, nonlinear, and quadratic attraction matrices before ACO tour construction.

# Training score: 5.795181757

# Search evaluation: 428

# Executable Algorithmic Primitives

def compute_attraction_score_matrix(distance_matrix: np.ndarray) -> np.ndarray:
    """Compute an attraction score matrix based on a modified polynomial function of distances; closer distances yield higher scores."""
    n = distance_matrix.shape[0]
    attraction_matrix = np.zeros_like(distance_matrix, dtype=np.float64)

    for i in range(n):
        for j in range(n):
            if distance_matrix[i, j] > 0:  # Avoid division by zero
                attraction_matrix[i, j] = 1 / (distance_matrix[i, j] ** 2)
    
    # Normalize attraction matrix row-wise to sum to 1 (optional, improve as needed)
    row_sums = attraction_matrix.sum(axis=1, keepdims=True)
    attraction_matrix = np.divide(attraction_matrix, row_sums, where=row_sums != 0)

    return attraction_matrix

def compute_complex_attraction_matrix(distance_matrix: np.ndarray) -> np.ndarray:
    """Computes an attraction score matrix using a combination of inverse distances and a sinusoidal function to introduce variability; closer distances yield higher scores with periodic attraction variations."""
    n = distance_matrix.shape[0]
    attraction_matrix = np.zeros((n, n))

    for i in range(n):
        for j in range(n):
            if i != j:
                inverse_distance = 1 / distance_matrix[i, j] if distance_matrix[i, j] != 0 else 0
                sinusoidal_factor = 1 + 0.5 * np.sin(np.pi * distance_matrix[i, j])
                attraction_matrix[i, j] = inverse_distance * sinusoidal_factor

    return attraction_matrix

def compute_quadratic_attraction_matrix(distance_matrix: np.ndarray) -> np.ndarray:
    """Computes an attraction score matrix using a quadratic function of inverse distances; closer distances yield exponentially higher scores."""
    # Initialize an attraction score matrix with the same shape as the distance matrix
    attraction_matrix = np.zeros_like(distance_matrix, dtype=np.float64)
    
    # Iterate through the distance matrix to compute attraction scores
    for i in range(distance_matrix.shape[0]):
        for j in range(distance_matrix.shape[1]):
            if distance_matrix[i, j] > 0:  # Avoid division by zero
                attraction_matrix[i, j] = 1 / (distance_matrix[i, j] ** 2)
            else:
                attraction_matrix[i, j] = np.inf  # Set to infinity if distance is zero
    
    return attraction_matrix

# Complete algorithm

import numpy as np
def heuristics(distance_matrix: np.ndarray) -> np.ndarray:
    """
    Design a heuristic matrix for ant colony optimization on TSP.

    Args:
        distance_matrix: Pairwise distance matrix of cities, shape (n, n).

    Returns:
        A finite non-negative heuristic matrix, shape (n, n). Larger values
        make an edge more attractive to ants.
    """
    n = distance_matrix.shape[0]
    heuristic_matrix = np.zeros((n, n))

    # Call the quadratic attraction matrix computation
    quadratic_attraction_matrix = compute_quadratic_attraction_matrix(distance_matrix)

    # Enhanced thresholding mechanism
    non_inf_quad_attract = quadratic_attraction_matrix[quadratic_attraction_matrix > 0]
    if non_inf_quad_attract.size > 0:
        threshold = np.percentile(non_inf_quad_attract, 75)
        quadratic_attraction_matrix[quadratic_attraction_matrix < threshold] = 0

    # Improved normalization process
    min_quad_attraction = np.nanmin(quadratic_attraction_matrix[quadratic_attraction_matrix > 0])
    max_quad_attraction = np.nanmax(quadratic_attraction_matrix)

    if max_quad_attraction > min_quad_attraction:
        heuristic_matrix = (quadratic_attraction_matrix - min_quad_attraction) / (max_quad_attraction - min_quad_attraction)

    # Call the complex attraction matrix computation
    attraction_matrix = compute_complex_attraction_matrix(distance_matrix)

    # Call the attraction score matrix computation
    attraction_score_matrix = compute_attraction_score_matrix(distance_matrix)

    heuristic_matrix = heuristic_matrix * attraction_matrix * attraction_score_matrix

    return heuristic_matrix
\end{lstlisting}

\begin{multicols}{2}
\subsection{Total Algorithm}
\label{app:pseudocode}

This subsection, we present the complete pseudo code of PACE, as shown in Algorithm \ref{alg:app-pace}.

\begin{algorithm*}[h!]
\caption{PACE search procedure}
\label{alg:app-pace}
\begin{algorithmic}[1]
\STATE Initialize population $P$ with valid complete algorithms.
\STATE Initialize EAP set $E$ and one $\mathrm{Beta}(1,1)$ posterior per EAP.
\FOR{$t=1$ to the search budget}
    \STATE Select parent algorithm(s) and operator $o_t$ from the fixed cycle.
    \IF{$o_t$ is \textit{Insertion} or \textit{Replacement}}
        \STATE Draw $z_e\sim\mathrm{Beta}(\alpha_e,\beta_e)$ for each eligible EAP.
        \STATE Select the focus EAP with largest draw; fix parent and EAP assignment.
    \ENDIF
    \STATE Request an offspring using the operator prompt.
    \FOR{$a=1$ to 3}
        \IF{the offspring violates the operator call-set contract}
            \STATE Request a repair with the same parent and EAP assignment.
        \ELSE
            \STATE Evaluate the offspring; \textbf{break}.
        \ENDIF
    \ENDFOR
    \IF{the child is valid and finite}
        \STATE Add the child to the history and update the population.
        \IF{the operator has focus EAP $e$}
            \IF{child score improves its fixed parent score}
                \STATE $\alpha_e\leftarrow\alpha_e+1$.
            \ELSE
                \STATE $\beta_e\leftarrow\beta_e+1$.
            \ENDIF
        \ENDIF
    \ENDIF
    \IF{the generation creates a strict best}
        \STATE Propose at most one deduplicated EAP through \textit{Extraction}.
    \ELSE
        \STATE Propose at most one deduplicated EAP through \textit{Generation}.
    \ENDIF
    \STATE Retain EAP source independently of population survival.
\ENDFOR
\RETURN best algorithm in the search history.
\end{algorithmic}
\end{algorithm*}

\section{Licenses}
\label{app:licenses}

\subsection{Implementation sources}

The programmatic AAD methods PACE, EoH, ReEvo, MCTS-AHD, HsEvo, and MLES are
executed through a common LLM4AD evaluation interface. We preserve each
method's search operators and method-specific population settings while
sharing the task evaluator and search budget. The classical and learned
baselines use their respective released implementations: Gymnasium provides
the control environments, Stable-Baselines3 provides PPO, and the released
DeepACO implementation provides the learned ACO baseline. Synthetic Euclidean
TSP data are generated by our scripts.

\subsection{Licenses}

Table~\ref{tab:app-licenses} reports the license attached to each software or
data resource.

\begin{table*}[h!]
\centering
\scriptsize
\setlength{\tabcolsep}{4pt}
\caption{Software and data resources used in the experiments. Copyright
notices and full license texts are retained in the released artifact. The
anonymous artifact URL is replaced by the archival URL after review.}
\begin{tabular}{p{0.18\linewidth}p{0.13\linewidth}p{0.20\linewidth}p{0.39\linewidth}}
\toprule
Resource & Type & License or notice & URL \\
\midrule
LLM4AD & Code & MIT License & \url{https://github.com/FeiLiu36/LLM4AD} \\
EoH & Code & MIT License & \url{https://github.com/FeiLiu36/EoH} \\
ReEvo & Code & MIT License & \url{https://github.com/ai4co/reevo} \\
HsEvo & Code & Available online & \url{https://github.com/datphamvn/HSEvo} \\
MCTS-AHD & Code & MIT License & \url{https://github.com/zz1358m/MCTS-AHD-master} \\
MLES & Code & MIT License & \url{https://github.com/QingL2000/MLES} \\
DeepACO & Code & MIT License & \url{https://github.com/henry-yeh/DeepACO} \\
\bottomrule
\end{tabular}
\label{tab:app-licenses}
\end{table*}

\end{multicols}



\end{document}